\documentclass[10pt,journal]{IEEEtran}

\usepackage[utf8]{inputenc} 
\usepackage[T1]{fontenc}

\usepackage[scaled]{helvet}
\usepackage{courier}

\usepackage{url}
\usepackage{ifthen}
\usepackage{cite}
\usepackage[cmex10]{amsmath} 
\usepackage{amssymb}
\usepackage{amsthm}
\usepackage{amsfonts}
\usepackage{mathtools} 
\usepackage{breqn} 
\usepackage{graphicx}
\usepackage{gensymb}
\usepackage{subcaption}
\usepackage{setspace}
\usepackage[inline]{enumitem}
\usepackage[linesnumbered,ruled,vlined]{algorithm2e}
\usepackage{pgfplots}
\usepackage{array}
\usepackage{textcomp}
\usepackage{stfloats}
\usepackage{titlesec}
\usepackage{verbatim}
\usepackage[hidelinks]{hyperref}
\usepackage{graphicx}
\usepackage{tabularx, multirow}
\usepackage{tikz}
\usetikzlibrary{shapes,arrows,positioning}
\usepackage{bbm, dsfont}
\pgfdeclareplotmark{mystar}{
    \node[star,star point ratio=2.25,minimum size=4pt,
          inner sep=0pt,draw=magenta,solid,fill=none] {};
}
\definecolor{col1}{RGB}{51,128,230}
\definecolor{col2}{RGB}{51,76,179}
\definecolor{col3}{RGB}{51,51,153}
\usepackage[left=0.625in, right=0.625in, top=0.75in, bottom=0.75in]{geometry}

\usetikzlibrary{arrows.meta, positioning, calc, shapes.geometric, fit, shadows, matrix, decorations.pathreplacing, patterns}

\newcommand{\x}{\boldsymbol{x}}

\newcommand{\F}{\mathbb{F}}

\newcommand*{\rv}[1]{\mathsf{#1}}

\newcommand*{\Rv}[1]{\boldsymbol{\mathsf{#1}}}

\newtheorem{theorem}{Theorem}

\newtheorem{lemma}[theorem]{Lemma}

\newtheorem{corollary}[theorem]{Corollary}

\newtheorem{remark}{Remark}

\newtheorem{definition}{Definition}

\definecolor{customgreen}{RGB}{77, 155, 74}
\definecolor{custombrown}{RGB}{166, 86, 40}
\definecolor{col1fade}{RGB}{214,230,250}
\definecolor{col3fade}{RGB}{174,174,214}

\newcommand\blfootnote[1]{%
  \begingroup
  \renewcommand\thefootnote{}\footnote{#1}%
  \addtocounter{footnote}{-1}%
  \endgroup
}
\allowdisplaybreaks[1]

\begin{document}
\title{On the Reliability of Information Retrieval \\From MDS Coded Data in DNA Storage}


\author{%
  \IEEEauthorblockN{Serge Kas Hanna} \\
  \IEEEauthorblockA{
C\^{o}te d’Azur University, CNRS, I3S Laboratory, France   \\
                    Email: serge.kas-hanna@cnrs.fr 
}
\vspace{-0.75cm}
}

\maketitle

\blfootnote{An earlier version of this paper was presented in part at the 2025 IEEE International Symposium on Information Theory (ISIT)~\cite{11195362}.}

\begin{abstract}
This work presents a theoretical analysis of the probability of successfully retrieving data encoded with MDS codes (e.g., Reed-Solomon codes) in DNA storage systems. We study this probability under independent and identically distributed (i.i.d.) substitution errors, focusing on a common code design strategy that combines inner and outer MDS codes.  Our analysis demonstrates how this probability depends on factors such as the total number of sequencing reads, their distribution across encoded sequences, the rates of the inner and outer codes, and the substitution error probabilities. These results provide actionable insights into optimizing DNA storage systems under reliability constraints, including determining the minimum number of sequencing reads needed for reliable data retrieval and identifying the optimal balance between the rates of inner and outer MDS~codes.
\end{abstract}

\section{Introduction}
The exponential growth of digital data has highlighted the limitations of conventional storage technologies in terms of scalability, energy efficiency, and longevity~\cite{rydning2022worldwide}. DNA-based data storage has emerged as a promising alternative due to its exceptional stability and ultra-high storage density~\cite{church2012next,grass2015robust,ceze2019molecular}. However, realizing efficient DNA storage systems requires addressing a variety of challenges, with reliability being a key concern~\cite{yazdi2015dna, shomorony2022information}. 
In this paradigm, digital data is encoded into short DNA sequences, which are written through synthesis as physical DNA strands, stored in molecular form (oligonucleotides), and read via sequencing. Reliability challenges arise from biochemical imperfections in this process, introducing single-base errors (i.e., deletions, insertions, and substitutions of nucleotides) that result in noisy reads of the underlying encoded sequences. Furthermore, biases introduced within the storage pipeline (e.g., PCR amplification bias) can lead to sequence dropouts, i.e., the loss of certain encoded sequences.

Abstractly, the DNA data storage channel is characterized by multiple sources of randomness~\cite{shomorony2022information,heckel2019characterization,meiser2020reading,gimpel2023digital}, including the stochastic nature of single-base errors; imbalances in physical coverage caused by biases; randomness in the order of reads in the sequencer output; and variability in sequencing coverage, i.e., different numbers of reads per encoded sequence due to biases and inherent randomness of sequencing. This work focuses on two main sources of randomness: substitution errors and variability in sequencing reads across the underlying encoded sequences, which can lead to both noisy reads and sequence dropouts. Error-correcting codes provide a robust solution to these reliability challenges~\cite{grass2015robust, blawat2016forward, erlich2017dna, yazdi2017portable, organick2018random, antkowiak2020low, chandak2020overcoming, press2020hedges, maarouf2022concatenated, welzel2023dna, 10619614, hanna2025gccodesystematicshort, 11154528}. A common design integrates inner and outer codes: the inner code adds redundancy within each encoded sequence to correct single-base errors, while the outer code introduces redundant sequences to recover from dropouts and correct residual errors from the inner code. Maximum distance separable (MDS) codes, such as Reed-Solomon codes~\cite{reed1960polynomial}, are a common choice in practice for both inner and outer codes due to their optimal error-correction properties~\cite{grass2015robust,erlich2017dna,organick2018random,antkowiak2020low, chandak2020overcoming,press2020hedges}.

In this work, we study the reliability of information retrieval from MDS coded data by characterizing the probability of successful retrieval as a function of the code and channel parameters. The work in~\cite{10750859} is most closely related to ours, where the authors derived bounds on the probability of successful retrieval for a coding scheme that uses an outer MDS code. In that study, the channel model accounted for a single source of randomness by modeling the sequencing step as a random sampling process with replacement, which reflects the variability in the number of reads across encoded sequences. The reliability analysis in~\cite{10750859} focused on the performance of the outer code under this sampling model, without considering single-base errors and inner codes. These results were later extended in~\cite{preuss2024sequencing,sokolovskii2024coding,cohen2024optimizing} to the case of combinatorial composites of DNA shortmers~\cite{preuss2024efficient}. Another related line of research, explored in~\cite{10750859,abraham2024covering,10619151}, focuses on constructing optimal outer codes that minimize the expected number of reads needed to retrieve parts or all of the information for noiseless channels. Also,~\cite{chandak2019improved} investigated the use of low-density parity-check (LDPC) codes to improve the read/write cost trade-offs in DNA storage systems.


\begin{figure*}
    \centering
    \resizebox{0.95\textwidth}{!}{%
\begin{tikzpicture}[
  font=\fontfamily{phv}\selectfont,
  >=Stealth,
  arrow/.style={-Stealth, thick, draw=gray!70},
bitbox/.style={
    draw=gray!90, fill=gray!5, rounded corners=1pt, thick,
    inner sep=4pt, font=\fontfamily{pcr}\selectfont\footnotesize
},
  rlabel/.style={font=\fontfamily{phv}\selectfont\scriptsize},
  bitbox_outer/.style={bitbox, fill=blue!15},
  bitbox_idx/.style={bitbox, fill=orange!15},
  bitbox_red/.style={bitbox, fill=red!15},
  bitbox_in/.style={bitbox, inner xsep=2pt} 
]


\coordinate (R1) at (0,0);
\coordinate (R2) at ([yshift=-6mm]R1);
\coordinate (RK) at ([yshift=-20mm]R1);  

\coordinate (RKp) at ([yshift=-6mm]RK);
\coordinate (RN)  at ([yshift=-16mm]RK);

\node[bitbox, anchor=west] (v11) at (R1) {01\ldots01};
\node[bitbox, right=0mm of v11] (v12) {10\ldots11};
\node[right=0mm of v12, font=\fontfamily{phv}\selectfont] (vell1) {$\cdots$};
\node[bitbox, right=0mm of vell1] (v1L) {00\ldots10};

\node[bitbox, anchor=west] (v21) at (R2) {11\ldots00};
\node[bitbox, right=0mm of v21] (v22) {01\ldots10};
\node[right=0mm of v22, font=\fontfamily{phv}\selectfont] (vell2) {$\cdots$};
\node[bitbox, right=0mm of vell2] (v2L) {11\ldots01};

\node[below=0mm of v21, font=\fontfamily{phv}\selectfont] {$\vdots$};
\node[below=0mm of v22, font=\fontfamily{phv}\selectfont] {$\vdots$};
\node[below=4mm of vell2, font=\fontfamily{phv}\selectfont] {$\cdots$};
\node[below=0mm of v2L, font=\fontfamily{phv}\selectfont] (anc1) {$\vdots$};

\node[bitbox, anchor=west] (vK1) at (RK) {10\ldots11};
\node[bitbox, right=0mm of vK1] (vK2) {00\ldots01};
\node[right=0mm of vK2, font=\fontfamily{phv}\selectfont] (vellK) {$\cdots$};
\node[bitbox, right=0mm of vellK] (vKL) {11\ldots10};

\node[bitbox_outer, anchor=west] (vKp1) at (RKp) {01\ldots10};
\node[bitbox_outer, right=0mm of vKp1] (vKp2) {10\ldots00};
\node[right=0mm of vKp2, font=\fontfamily{phv}\selectfont] (vellKp) {$\cdots$};
\node[bitbox_outer, right=0mm of vellKp] (vKpL) {00\ldots11};

\node[below=-2mm of vKp1, font=\fontfamily{phv}\selectfont] {$\vdots$};
\node[below=-2mm of vKp2, font=\fontfamily{phv}\selectfont] {$\vdots$};
\node[below=2mm of vellKp, font=\fontfamily{phv}\selectfont] {$\cdots$};
\node[below=-2mm of vKpL, font=\fontfamily{phv}\selectfont] {$\vdots$};

\node[bitbox_outer, anchor=west] (vN1) at (RN) {11\ldots01};
\node[bitbox_outer, right=0mm of vN1] (vN2) {00\ldots10};
\node[right=0mm of vN2, font=\fontfamily{phv}\selectfont] (vellN) {$\cdots$};
\node[bitbox_outer, right=0mm of vellN] (vNL) {10\ldots11};

\draw [decorate, decoration={brace, amplitude=5pt, mirror}] 
        ($(v11.north west)+(-0.1,0)$) -- ($(vK1.south west)+(-0.1,0)$) 
        node [midway, left=6pt, font=\fontfamily{phv}\selectfont\footnotesize] {$K$};

\draw [decorate, decoration={brace, amplitude=5pt, mirror}] 
        ($(vKp1.north west)+(-0.1,0)$) -- ($(vN1.south west)+(-0.1,0)$) 
        node [midway, left=6pt, font=\fontfamily{phv}\selectfont\footnotesize] {$N-K$};

\draw[<->, black]
  ([yshift=1.0mm]v11.north west) -- ([yshift=1.0mm]v11.north east)
  node[midway, above, font=\fontfamily{phv}\selectfont\footnotesize] {$M$};
\draw[<->, black]
  ([yshift=1.0mm]v1L.north west) -- ([yshift=1.0mm]v1L.north east)
  node[midway, above, font=\fontfamily{phv}\selectfont\footnotesize] {$M$};

\draw[draw=gray!90, densely dashed, thick, rounded corners=3pt] 
      (-0.95,1.2) rectangle (5.3,-2.3);

\draw[draw=gray!90, densely dashed, thick, rounded corners=3pt] 
      (-1.45,1.6) rectangle (5.45,-4.2);
      
\node[above] at (2.175,1.12) {\fontfamily{phv}\selectfont\scriptsize\color{gray} (1) Fragmentation};
\node[above] at (2,1.55) {\fontfamily{phv}\selectfont\scriptsize\color{gray} (2) Outer $(N,K)$ MDS code over $\mathbb{F}_{2^M}$ (inter-sequence)};

\draw [arrow, densely dotted, black, thin] ($(v11.north west)+(0.09,0)$ ) -- ($(vN1.south west)+(0.09,-0.3)$);
\draw [arrow, densely dotted, black, thin] ($(v1L.north west)+(0.09,0)$ ) -- ($(vNL.south west)+(0.09,-0.3)$);
\draw[<->, black]
  ([yshift=5.5mm]v11.north west) -- ([yshift=5.5mm]v1L.north east)
  node[midway, above, font=\fontfamily{phv}\selectfont\footnotesize] {$k$};


\coordinate (S1)  at ([xshift=12mm]v1L.east);
\coordinate (S2)  at ([xshift=12mm]v2L.east);
\coordinate (SK2) at ([xshift=12mm]vKL.east);
\coordinate (SKp2) at ([xshift=12mm]vKpL.east);
\coordinate (SN2) at ([xshift=12mm]vNL.east);

\tikzset{
  bitbox_in_reg/.style={bitbox_in},
  bitbox_in_out/.style={bitbox_outer, inner xsep=2pt},
  bitbox_red_in/.style={bitbox_red, inner xsep=2pt}
}

\node[bitbox_in_reg, anchor=west] (w12) at (S1) {0\ldots1};
\node[bitbox_in_reg, right=0mm of w12] (w13) {1\ldots0};
\node[right=0mm of w13, font=\fontfamily{phv}\selectfont] (w1dotsA) {$\cdots$};
\node[bitbox_in_reg, right=0mm of w1dotsA] (w14) {0\ldots0};
\node[bitbox_red_in, right=0mm of w14] (w15r) {1\ldots1};
\node[right=0mm of w15r, font=\fontfamily{phv}\selectfont] (w1dotsR) {$\cdots$};
\node[bitbox_red_in, right=0mm of w1dotsR] (w1Lr) {0\ldots1};

\node[bitbox_in_reg, anchor=west] (w22) at (S2) {1\ldots1};
\node[bitbox_in_reg, right=0mm of w22] (w23) {0\ldots0};
\node[right=0mm of w23, font=\fontfamily{phv}\selectfont] (w2dotsA) {$\cdots$};
\node[bitbox_in_reg, right=0mm of w2dotsA] (w24) {1\ldots1};
\node[bitbox_red_in, right=0mm of w24] (w25r) {0\ldots1};
\node[right=0mm of w25r, font=\fontfamily{phv}\selectfont] (w2dotsR) {$\cdots$};
\node[bitbox_red_in, right=0mm of w2dotsR] (w2Lr) {1\ldots1};

\node[below=0mm of w22,  font=\fontfamily{phv}\selectfont] (anc2) {$\vdots$};
\node[below=0mm of w23,  font=\fontfamily{phv}\selectfont] {$\vdots$};
\node[below=4mm of w2dotsA, font=\fontfamily{phv}\selectfont] {$\cdots$};
\node[below=0mm of w24, font=\fontfamily{phv}\selectfont] {$\vdots$};
\node[below=0mm of w25r, font=\fontfamily{phv}\selectfont] {$\vdots$};
\node[below=4mm of w2dotsR, font=\fontfamily{phv}\selectfont] {$\cdots$};
\node[below=0mm of w2Lr, font=\fontfamily{phv}\selectfont] (anc3) {$\vdots$};

\node[bitbox_in_reg, anchor=west] (wK2) at (SK2) {1\ldots0};
\node[bitbox_in_reg, right=0mm of wK2] (wK3) {0\ldots1};
\node[right=0mm of wK3, font=\fontfamily{phv}\selectfont] (wKdotsA) {$\cdots$};
\node[bitbox_in_reg, right=0mm of wKdotsA] (wK4) {1\ldots0};
\node[bitbox_red_in, right=0mm of wK4] (wK5r) {0\ldots0};
\node[right=0mm of wK5r, font=\fontfamily{phv}\selectfont] (wKdotsR) {$\cdots$};
\node[bitbox_red_in, right=0mm of wKdotsR] (wKLr) {1\ldots0};

\node[bitbox_in_out, anchor=west] (wKp2) at (SKp2) {0\ldots1};
\node[bitbox_in_out, right=0mm of wKp2] (wKp3) {1\ldots1};
\node[right=0mm of wKp3, font=\fontfamily{phv}\selectfont] (wKpdotsA) {$\cdots$};
\node[bitbox_in_out, right=0mm of wKpdotsA] (wKp4) {0\ldots1};
\node[bitbox_red_in, right=0mm of wKp4] (wKp5r) {1\ldots0};
\node[right=0mm of wKp5r, font=\fontfamily{phv}\selectfont] (wKpdotsR) {$\cdots$};
\node[bitbox_red_in, right=0mm of wKpdotsR] (wKpLr) {0\ldots1};

\node[below=2mm of wKpdotsA, font=\fontfamily{phv}\selectfont] {$\cdots$};
\node[below=-2mm of wKp4, font=\fontfamily{phv}\selectfont] {$\vdots$};
\node[below=-2mm of wKp5r, font=\fontfamily{phv}\selectfont] {$\vdots$};
\node[below=2mm of wKpdotsR, font=\fontfamily{phv}\selectfont] {$\cdots$};
\node[below=-2mm of wKpLr, font=\fontfamily{phv}\selectfont] {$\vdots$};

\node[bitbox_in_out, anchor=west] (wN2) at (SN2) {1\ldots0};
\node[bitbox_in_out, right=0mm of wN2] (wN3) {0\ldots1};
\node[right=0mm of wN3, font=\fontfamily{phv}\selectfont] (wNdotsA) {$\cdots$};
\node[bitbox_in_out, right=0mm of wNdotsA] (wN4) {1\ldots1};
\node[bitbox_red_in, right=0mm of wN4] (wN5r) {0\ldots0};
\node[right=0mm of wN5r, font=\fontfamily{phv}\selectfont] (wNdotsR) {$\cdots$};
\node[bitbox_red_in, right=0mm of wNdotsR] (wNLr) {1\ldots0};

\draw [decorate, decoration={brace, amplitude=5pt}] 
        ($(w15r.north west)+(0,0.4)$) -- ($(w1Lr.north east)+(0,0.4)$) 
        node [midway, above=3.5pt, font=\fontfamily{phv}\selectfont\footnotesize] {$n'-k'=\frac{n-k}{m}$};

\draw [decorate, decoration={brace, amplitude=5pt}] 
        ($(w12.north west)+(0,0.4)$) -- ($(w14.north east)+(0,0.4)$) 
        node [midway, above=3.5pt, font=\fontfamily{phv}\selectfont\footnotesize] {$k'=\frac{k}{m}$};
        

   \draw[draw=gray!90, densely dashed, thick, rounded corners=3pt] 
      (6.25,1.4) rectangle (13.0,-4.2);

\node[above] at (9.6,1.32) {\fontfamily{phv}\selectfont\scriptsize\color{gray} (3) Inner $(n',k')$ MDS code over $\mathbb{F}_{2^m}$ (intra-sequence)};

\draw[<->, black]
  ([yshift=1mm]w12.north west) -- ([yshift=1mm]w12.north east)
  node[midway, above, font=\fontfamily{phv}\selectfont\footnotesize] {$m$};


\draw [arrow, densely dotted, black, thin] ($(w12.south west)+(0,0.08)$ ) -- ($(w1Lr.south east)+(0.25,0.08)$);
\draw [arrow, densely dotted, black, thin] ($(wN2.south west)+(0,0.08)$ ) -- ($(wNLr.south east)+(0.25,0.08)$);


\coordinate (Q1)  at ([xshift=11.5mm]w1Lr.east);
\coordinate (Q2)  at ([xshift=11.5mm]w2Lr.east);
\coordinate (QK2) at ([xshift=11.5mm]wKLr.east);
\coordinate (QKp2) at ([xshift=11.5mm]wKpLr.east);
\coordinate (QN3) at ([xshift=11.5mm]wNLr.east);

\tikzset{
  dna_gray_wide/.style={bitbox_in, inner xsep=4pt},
  dna_blue_wide/.style={bitbox_outer, inner xsep=4pt},
  dna_red_small/.style={bitbox_red, inner xsep=4pt}
}

\node[dna_gray_wide, anchor=west] (z1data) at (Q1) {TCGT\ldots\ldots TGAC};
\node[dna_red_small, right=0mm of z1data] (z1red) {GC\ldots AT};

\node[dna_gray_wide, anchor=west] (z2data) at (Q2) {GTCA\ldots\ldots ACGT};
\node[dna_red_small, right=0mm of z2data] (z2red) {AT\ldots CG};

\node[below=0mm of z2data, font=\fontfamily{phv}\selectfont] (anc4) {$\vdots$};
\node[below=0mm of z2red, font=\fontfamily{phv}\selectfont] {$\vdots$};

\node[dna_gray_wide, anchor=west] (zKdata) at (QK2) {CGTA\ldots\ldots GTAC};
\node[dna_red_small, right=0mm of zKdata] (zKred) {TA\ldots CA};

\node[dna_blue_wide, anchor=west] (zKpdata) at (QKp2) {TGAC\ldots\ldots CGTG};
\node[dna_red_small, right=0mm of zKpdata] (zKpred) {CG\ldots AT};

\node[below=-2mm of zKpdata, font=\fontfamily{phv}\selectfont] {$\vdots$};
\node[below=-2mm of zKpred, font=\fontfamily{phv}\selectfont] {$\vdots$};

\node[dna_blue_wide, anchor=west] (zNdata) at (QN3) {GACT\ldots\ldots ATCG};
\node[dna_red_small, right=0mm of zNdata] (zNred) {TC\ldots AC};

\draw[draw=gray!90, densely dashed, thick, rounded corners=3pt] 
      (13.79,1.4) rectangle (18.1,-4.2);

\node[above] at (15.87,1.32) {\fontfamily{phv}\selectfont\scriptsize\color{gray} (4) DNA Mapping};

\draw [<->, black] 
        ([yshift=5.5mm]z1data.north west) -- ([yshift=5.5mm]z1data.north east) 
        node [midway, above, font=\fontfamily{phv}\selectfont\footnotesize] {$\frac{k}{2}$};


\coordinate (LegendCenter) at ($(v11.west)!0.5!(z1red.east)$);
\coordinate (LegendBase)   at ($(LegendCenter)+(0,2.45)$);
\coordinate (L1) at ($(LegendBase)+(-9,0)$);
\coordinate (L2) at ($(LegendBase)+(-5.1,0)$);
\coordinate (L3) at ($(LegendBase)+(2,0)$);

\node[bitbox, anchor=west] (leg1) at (L1) {\phantom{0\ldots0}};
\node[anchor=west, font=\fontfamily{phv}\selectfont\scriptsize] at ([xshift=1mm]leg1.east)
  {Information};

\node[bitbox_outer, anchor=west] (leg2) at (L2) {\phantom{0\ldots0}};
\node[anchor=west, font=\fontfamily{phv}\selectfont\scriptsize] at ([xshift=1mm]leg2.east)
  {Outer code (inter-sequence) redundancy};

\node[bitbox_red, anchor=west] (leg3) at (L3) {\phantom{0\ldots0}};
\node[anchor=west, font=\fontfamily{phv}\selectfont\scriptsize] at ([xshift=1mm]leg3.east)
  {Inner code (intra-sequence) redundancy};
  
\draw [<->,black] 
        ([yshift=5.5mm]z1red.north west) -- ([yshift=5.5mm]z1red.north east) 
        node [midway, above, font=\fontfamily{phv}\selectfont\footnotesize] {$\frac{n-k}{2}$};


\draw[arrow, gray, very thick] ($(anc1)!0.5!(anc2)$) -- ++(0.5,0);
\draw[arrow, gray, very thick] ($(anc3)!0.5!(anc4)+(-0.525,0)$) -- ++(0.5,0);
\end{tikzpicture}
    }
\caption{Schematic illustration of the encoding process, assuming systematic inner and outer MDS codes. (1)~The input data is partitioned into $K$ fragments, each of length $k$ bits.
(2)~An outer $(N,K)$ MDS code over $\mathbb{F}_{2^M}$ is applied across fragments to introduce inter-sequence redundancy, where each code symbol contains $M$ bits.
(3)~Each fragment is then individually encoded using an inner \mbox{$(n'=\frac{n}{m},k'=\frac{k}{m})$} MDS code over $\mathbb{F}_{2^m}$ to introduce intra-sequence redundancy, where each code symbol contains $m$ bits.
(4)~The resulting binary sequences are mapped to quaternary DNA sequences according to $00 \leftrightarrow \mathsf{A}$, $01 \leftrightarrow \mathsf{T}$, $10 \leftrightarrow \mathsf{C}$, $11 \leftrightarrow \mathsf{G}$.}
\label{fig:scheme}
\vspace{-0.4cm}
\end{figure*}

Most existing theoretical analyses of data retrieval reliability have concentrated on outer MDS codes and sequencing randomness, while assuming error-free reads and neglecting the role of inner codes. As a result, these studies do not capture the impact of residual single-base errors on the outer decoder, nor the interplay between the inner and outer codes. These works have nonetheless provided valuable insights, including interesting connections to classical probability problems such as the coupon collector and double Dixie cup problems. In this paper, we extend this line of research by providing a theoretical analysis of the interaction between inner and outer codes while accounting for randomness from both sequencing and error processes. Guided by recent experimental evidence that substitution errors tend to occur independently and that coverage bias is substantial in practice~\cite{gimpel2023digital}, we adopt an i.i.d. substitution model for single-base errors and allow encoded sequences to be sampled with unequal probabilities in the sequencing process.

Our main contribution is a theoretical framework presented in Section~\ref{sec3}, which evaluates the analytical probability of successful retrieval in terms of the channel and code parameters. This framework provides insights into trade-offs between system parameters, enabling the optimization of sequencing and synthesis costs. In Section~\ref{opt}, we illustrate this through numerical examples addressing two optimization problems under reliability constraints: minimizing number of reads and maximizing information density. In Section~\ref{sec5}, we discuss how our framework can be applied to other inner/outer codes and to additional error types, such as deletions and insertions.

\section{Preliminaries} \label{sec2}

\subsection{Notation} \label{sec2a}
We use bold letters for vectors, sans-serif letters for random variables, and calligraphic letters for sets; e.g., $\boldsymbol{x}$, $\rv{X}$, $\Rv{X}$, and $\mathcal{X}$ represent a vector, a random variable, a random vector, and a set, respectively. For integers $n$, $i$, and $j\geq i$, we define \mbox{$[n]\triangleq \{1,2,\ldots,n\}$} and \mbox{$[i,j]\triangleq\{i,i+1,\ldots,j\}$}. Superscripts index vectors, e.g., $\x^i$, and subscripts index elements within a vector, e.g.,~$x_i$. For a vector $\boldsymbol{x}$, $\boldsymbol{x}_{[i,j]}$ represents the subvector containing the  elements of $\boldsymbol{x}$ indexed by $[i,j]$. We write a sequence of vectors $\boldsymbol{x}^1,\boldsymbol{x}^2, \ldots, \boldsymbol{x}^n$ compactly as $(\boldsymbol{x}^i)_{i=1}^n$. We use $\|\boldsymbol{x}\|_0$ to denote to the number of non-zero elements in~$\boldsymbol{x}$. Let $\mathbb{N}=\{0,1,2,\ldots\}$ be the set of natural numbers and $\F_q$ be the Galois field of size $q$. We denote the indicator function by \( \mathds{1}_{\{\text{condition}\}} \in \{0,1\} \), which equals $1$ when the condition is true and $0$ otherwise. The probability of an event is denoted by $\Pr(\rv{X} = x)$, where we sometimes omit the random variable, e.g., $\Pr(x)$, when it is clear from the
context. The probability mass function (PMF) and cumulative distribution function (CDF) of a binomial random variable  $\rv{X} \sim \text{Binomial}(n, p)$ are denoted by \mbox{$f(x; n, p) \triangleq \Pr(\rv{X}=x)= \binom{n}{x} p^x (1-p)^{n-x}$} and \mbox{$F(x; n, p) \triangleq \Pr(\rv{X}\leq x) = \sum_{j=0}^x f(j; n, p)$}, respectively. The CDF of a standard Gaussian distribution is denoted by \mbox{$\Phi (x)= \frac{1}{\sqrt{2\pi}} \int_{-\infty}^x e^{-\frac{z^2}{2}} \, dz$}, where $x \in \mathbb{R}$.

\subsection{Encoding}

We consider a coding scheme that encodes binary information in the form of a collection of DNA sequences using inner and outer MDS codes. The encoding process is illustrated in Fig.~\ref{fig:scheme} and described formally next. The input binary information, denoted by $\boldsymbol{u}$, is first partitioned into $K$ non-overlapping sequences $(\boldsymbol{u}^j)_{j=1}^K$, each of length $k$ bits, where \mbox{$\boldsymbol{u}^j \triangleq \boldsymbol{u}_{[1+(j-1)k, jk]}\in \mathbb{F}_2^k$}. These $K$ information sequences are encoded using an outer $(N,K)$ MDS code, resulting in $(\boldsymbol{w}^j)_{j=1}^N$, where \mbox{$\boldsymbol{w}^j\in \mathbb{F}_2^k$}. The outer MDS code operates ``column-wise'' on symbols from different sequences over $\mathbb{F}_{2^M}$, with $2^M\geq N$, where each $\boldsymbol{w}^j$ is represented as a sequence of $M$-bit symbols. Subsequently, each $\boldsymbol{w}^j \in \mathbb{F}_2^k$ is encoded into $\boldsymbol{x}^j \in \mathbb{F}_2^n$ using an inner \mbox{$(n'=\frac{n}{m},k'=\frac{k}{m})$}~MDS code that operates on $m$-bit symbols over $\mathbb{F}_{2^m}$, with $2^m\geq n'$, and $n'-k'=2t$. The combination of inner and outer MDS codes yields $N$ encoded sequences $(\boldsymbol{x}^j)_{j=1}^N$, where $\boldsymbol{x}^j \in \mathbb{F}_2^n$.

To complete the encoding process, each binary sequence  $\boldsymbol{x}^j$ is mapped into a DNA representation $\tilde{\boldsymbol{x}}^j$ via a one-to-one mapping that replaces every two bits with a symbol from $\Sigma \triangleq \{\mathsf{A,C,G,T}\}$, e.g., $00 \leftrightarrow \mathsf{A}$, $01 \leftrightarrow \mathsf{T}$, $10 \leftrightarrow \mathsf{C}$, $11 \leftrightarrow \mathsf{G}$. As a result, the encoder outputs $N$ DNA sequences $(\tilde{\boldsymbol{x}}^j)_{j=1}^N$, where $\tilde{\boldsymbol{x}}^j \in \Sigma^{n/2}$. The resulting information density, defined as the amount of information bits encoded per DNA nucleotide, is $\Delta \triangleq 2 \times \rho_{\text{in}} \times \rho_{\text{out}}$, where $\rho_{\text{in}} = \frac{k}{n}$ and $\rho_{\text{out}} = \frac{K}{N}$ are the code rates of the inner and outer MDS codes, respectively. For simplicity, we assume all parameters are divisible as needed so that the integer-valued quantities (e.g., $k/m$, $n/m$, and $n/2$) are well-defined.

\subsection{Channel Model} \label{channel}

Given the $N$ encoded DNA sequences $(\tilde{\boldsymbol{x}}^j )_{j=1}^N$ as input, we consider a channel that outputs $\rv{R}_j \in \mathbb{N}$  noisy copies of each sequence \mbox{$\tilde{\boldsymbol{x}}^j \in \Sigma^{n/2}$}. The noise is introduced through random nucleotide substitutions, independently across all sequences and their copies. The channel output is represented as $((\tilde{\Rv{Y}}^{j,\ell} )_{\ell=1}^{\rv{R}_j})_{j=1}^N$, where \mbox{$\tilde{\Rv{Y}}^{j,\ell} \in \Sigma^{n/2}$} denotes the $\ell^{\text{th}}$ noisy copy of $\tilde{\boldsymbol{x}}^j$. Throughout this paper, we use the term {\em read profile} to refer to the random vector \mbox{$\Rv{R}=(\rv{R}_1,\ldots,\rv{R}_N)$} and its realization \mbox{$\boldsymbol{r}=(r_1,\ldots,r_N)$}, which specify the number of sequencing reads (noisy copies) per encoded sequence.

Similar to previous works, we model the sequencer output as performing $R_{\text{\normalfont all}}$ independent draws (with replacement) from a probability distribution $\boldsymbol{p} = (p_1, \ldots, p_N)$, with $0\leq p_j\leq 1$ and $\sum_{j=1}^N p_j=1$. Here, $R_{\text{\normalfont all}}\in \mathbb{N}$ represents the total number of sequencing reads, and $p_j$ reflects the probability of reading sequence~$\tilde{\boldsymbol{x}}^j$, which is related in practice to factors such as its physical coverage. Under this model, the read profile follows a multinomial distribution with parameters $R_{\text{\normalfont all}}$ and~$\boldsymbol{p}$, and hence
\begin{equation} \label{eqM}
\Pr(\boldsymbol{r})=\Pr(r_1,\ldots,r_N)=\binom{R_{\text{\normalfont all}}}{r_1,\ldots,r_N} \prod_{j=1}^N p_j^{r_j}.
\end{equation}
We also consider a closely related model in which $\rv{R}_1,\ldots,\rv{R}_N$ are independent Poisson random variables with \mbox{$\rv{R}_j \sim \text{Pois}(\lambda p_j)$} for $j\in [N]$, and thus
\begin{equation} \label{eqP}
\Pr(\boldsymbol{r}) = \Pr(r_1,\ldots,r_N) = \prod_{j=1}^N \frac{(\lambda p_j)^{r_j}e^{-\lambda p_j}}{r_j!}.
\end{equation}
The connection between both models stems from the well-known \emph{Poissonization} phenomenon applied to multinomials. Specifically, if we replace the fixed total number of reads $R_{\text{\normalfont all}}$ in the multinomial model with a Poisson-distributed random variable with mean $\lambda$, i.e., $\rv{R}_{\text{\normalfont all}} \sim \text{Pois}(\lambda)$, and marginalize over $\rv{R}_{\text{\normalfont all}}$ in~\eqref{eqM}, we obtain the same distribution in~\eqref{eqP}. Furthermore, for large fixed $R_{\text{\normalfont all}}$ and small sampling probabilities $p_1, \ldots, p_N$, a common approximation of~\eqref{eqM} follows from substituting $\lambda$ with $R_{\text{\normalfont all}}$ in~\eqref{eqP}.

Finally, motivated by recent experimental results in DNA storage showing that the assumption of error independence is generally valid in practice for substitution errors~\cite{gimpel2023digital}, we model the noise in each read as i.i.d. nucleotide substitutions occurring with probability~$\epsilon>0$. More precisely, we consider a memoryless quaternary symmetric channel (QSC) with error probability $\epsilon>0$, such that for a given input sequence $\tilde{\boldsymbol{x}}^j \in \Sigma^{n/2}$, read profile $\boldsymbol{r}=(r_1,\ldots,r_N)$, and any \mbox{$j\in [N]$}, $\ell \in [r_j]$, $i \in [n/2]$, and $\tilde{y} \in \Sigma$, we have 
\begin{equation} \label{QSC}
\Pr(\tilde{\rv{Y}}^{j,\ell}_i= \tilde{y} \mid \tilde{x}^j_i) =
\begin{cases} 
1 - \epsilon, & \text{if }  \tilde{y} = \tilde{x}^j_i, \\ 
\frac{\epsilon}{3}, & \text{if } \tilde{y} \neq \tilde{x}^j_i.
\end{cases}
\end{equation}

\subsection{Decoding} \label{deco}
The first step in the decoding process generally involves clustering the sequencer output to identify and group the reads belonging to the same underlying encoded sequence. The channel model considered in this work does not account for randomness in the order of the reads, so we assume the channel output $((\tilde{\Rv{Y}}^{j,\ell} )_{\ell=1}^{\rv{R}_j})_{j=1}^N$ is ordered such that the reads corresponding to each encoded sequence are identifiable, making clustering implicit and error-free. This assumption also implies that the read profile $\boldsymbol{r}=(r_1,\ldots,r_N)$ is known at the decoder. We defer the analysis of the effect of clustering on the retrieval process to future work.

For each sequence index $j\in [N]$, the noisy reads $(\tilde{\Rv{Y}}^{j,\ell} )_{\ell=1}^{r_j}$ are aggregated to produce a single {\em consensus} sequence \mbox{$\tilde{\Rv{C}}^j\in \Sigma^{n/2}$} via base-by-base majority voting over $\Sigma = \{\mathsf{A,C,G,T}\}$. Specifically, for each position $i\in [n/2]$, the consensus nucleotide $\tilde{\rv{C}}^j_i\in \Sigma$ is chosen as the most frequent among $\tilde{\rv{Y}}^{j,1},\ldots, \tilde{\rv{Y}}^{j,r_j}$, with ties broken uniformly at random. The consensus sequences $(\tilde{\Rv{C}}^{j} )_{j=1}^{N}$  are then mapped to their binary representations $(\Rv{C}^{j} )_{j=1}^{N}$, where $\Rv{C}^{j} \in \mathbb{F}_2^n$, by applying the inverse of the mapping used during encoding. Each consensus sequence $\Rv{C}^j$ is subsequently decoded using the inner \mbox{$(n'=\frac{n}{m},k'=\frac{k}{m})$}~MDS code, yielding $(\hat{\Rv{W}}^j)_{j=1}^N$, where $\hat{\Rv{W}}^j \in \mathbb{F}_2^k$. The outer $(N,K)$ MDS code over $\mathbb{F}_{2^M}$ then recovers $(\hat{\Rv{U}}^j)_{j=1}^K$ from $(\hat{\Rv{W}}^j)_{j=1}^N$, where $\hat{\Rv{U}}^j \in \mathbb{F}_2^k$. Finally, the information sequences $(\hat{\Rv{U}}^j)_{j=1}^K$ are concatenated to form the final estimate $\hat{\Rv{U}}$ of the original information $\boldsymbol{u}$. We define the probability of successful retrieval as $P_{\text{succ}}\triangleq \Pr(\hat{\Rv{U}} = \boldsymbol{u})$, which depends on the parameters of the inner and outer MDS codes, as well as the channel parameters.

\begin{remark}[Rate-reliability trade-off in reconstruction] There are multiple approaches to reconstructing an individual MDS-coded sequence from multiple noisy reads. For instance, the inner decoding can be performed directly on noisy reads before forming a consensus, or the consensus itself could be performed at the sequence level rather than on a base-by-base basis. In this work, we focus on the case where base-by-base consensus is performed as a preliminary step before applying the inner MDS decoder. We conjecture that this approach optimizes the trade-off between the rate of the inner MDS code and the reliability of reconstruction in the presence of i.i.d. substitution~errors. \end{remark}

\begin{figure*}[h!]


\begin{equation} \label{eqP0}
\Pr\left(\rv{S}_N{(\boldsymbol{r})}\geq K \right) = \sum_{\substack{(\mathcal{A}, \mathcal{B}) \subseteq \mathcal{N}}} \bigg(  \prod_{j \in \mathcal{A}} \alpha_{(r_{j})} \bigg) \bigg( \prod_{j \in \mathcal{B}} \beta_{(r_j)} \bigg)  \bigg( \prod_{j \notin \mathcal{A} \cup \mathcal{B}} \gamma_{(r_j)} \bigg), \quad \mathcal{N} \triangleq \{(\mathcal{A},\mathcal{B}) : \mathcal{A},\mathcal{B} \subseteq [N],\mathcal{A} \cap \mathcal{B} = \varnothing,  |\mathcal{A}| - |\mathcal{B}|\geq K \}.
\end{equation}

   \medskip
\hrulefill\par

\vspace{-0.4cm}
       \end{figure*}

\section{Reliability of Data Retrieval} \label{sec3}
In this section, we evaluate the theoretical probability of successful data retrieval by examining key aspects of the decoding process. This includes analyzing the residual error rate in the consensus sequence, the probabilities of success, miscorrection, and failure when decoding individual sequences using the inner MDS code, and the probability of successful decoding of the outer MDS code. These probabilities are influenced by various factors, such as the QSC error rate, the number of reads and their distribution among the encoded sequences, and the parameters of the inner and outer MDS codes. Formal proofs of the presented results are provided in the Appendix.

\subsection{Consensus and Inner MDS Code} \label{sec3a}
Given $r\in \mathbb{N}$ noisy reads of a DNA sequence $\tilde{\boldsymbol{x}}\in \Sigma^{n/2}$, a preliminary decoding step, as outlined in Section~\ref{sec2}, involves deriving a consensus sequence $\tilde{\Rv{C}}$ via majority voting. This step can correct single-base substitution errors by leveraging redundancy in the reads, with its effectiveness depending on the QSC error rate $\epsilon$ and the number of reads $r$. Consequently, the error rate in the consensus sequence, denoted by~$\epsilon_{(r)}$, satisfies $\epsilon_{(r)}\leq \epsilon$. Moreover, since the QSC introduces i.i.d. errors and the consensus mechanism processes each nucleotide position independently, the errors in $\tilde{\Rv{C}}$ also remain i.i.d., where $\epsilon_{(r)}=\Pr(\tilde{\rv{C}}_i \neq \tilde{x}_i)$ for any $i \in [n/2]$. In Lemma~\ref{lemm1}, we establish the exact relationship between $\epsilon$ and $\epsilon_{(r)}$ as a function of $r$.

The inner MDS decoder is then applied to the consensus sequence, and its performance depends directly on the post-consensus error rate $\epsilon_{(r)}$. Since the inner MDS code addresses symbol-level errors, where each symbol corresponds to $m$ consecutive bits (i.e., $m/2$ nucleotides), we also provide the corresponding {\em symbol} error rate in Lemma~\ref{lemm1}, denoted by $\epsilon'_{(r)}$. The value of $\epsilon'_{(r)}$ represents the {\em effective} error rate relevant to the inner MDS code, determining its ability to handle residual errors that remain after the consensus process.
\begin{lemma} \label{lemm1}
    For a given number of reads $r\in \mathbb{N}$, the post-consensus nucleotide error rate is given by
    $$\epsilon_{(r)} = 1-\sum_{\boldsymbol{\kappa}\in \mathcal{K}_{(r)}} \frac{1}{\omega_{(\boldsymbol{\kappa})}} \binom{r}{\kappa_1,\kappa_2,\kappa_3,\kappa_4}(1-\epsilon)^{\kappa_1} \left(\frac{\epsilon}{3} \right)^{r-\kappa_1},$$
    and the inner code symbol error rate is $\epsilon'_{(r)} = 1 - \left(1- \epsilon_{(r)}\right)^{\frac{m}{2}}$, where $\omega_{(\boldsymbol{\kappa})} \triangleq  \sum_{i=1}^4 \mathds{1}_{\{\kappa_i = \kappa_1\}}$ and $$ \mathcal{K}_{(r)} \triangleq \bigg\{ \boldsymbol{\kappa} \in \mathbb{N}^4 \ : \ \sum_{i=1}^{4} \kappa_i = r, \kappa_1 \geq  \kappa_2,  \kappa_3,  \kappa_4 \bigg\}. $$
\end{lemma}
The decoding of the inner MDS code can result in one of three outcomes:~a successful decoding (recovering the correct sequence information), a miscorrection (producing an incorrect result), or a decoding failure (acknowledging an inability to decode). The probabilities of these three outcomes are formally defined as follows.

\begin{definition}[Inner-decoder outcome probabilities]
\label{def:abg}
Given \mbox{$r\in \mathbb{N}$} noisy reads of a sequence $\tilde{\boldsymbol{x}}$, we define the inner-decoder probabilities of success, miscorrection, and failure by
$$\alpha_{(r)} \triangleq \Pr(\hat{\Rv{W}} = \boldsymbol{w}), \beta_{(r)} \triangleq \Pr(\hat{\Rv{W}} \neq \boldsymbol{w}), \gamma_{(r)}\triangleq \Pr(\hat{\Rv{W}} = \varnothing), $$
respectively, with $\gamma_{(0)}=1$ and \mbox{$\alpha_{(r)} + \beta_{(r)} + \gamma_{(r)} = 1$}.
\end{definition}

In Lemma~\ref{lemma2}, we provide exact expressions for these probabilities as functions of $r$ and the inner code parameters.

\begin{lemma}
\label{lemma2}
Consider the inner \mbox{$(n'=\frac{n}{m},k'=\frac{k}{m})$} MDS code over $\mathbb{F}_{2^m}$. Under bounded-distance decoding (BDD) with radius $t=(n'-k')/2$, we have
   \begin{align*}
   \alpha_{(r)} &= F_{\text{\normalfont}}(t; n', \epsilon'_{(r)}), \\
   \beta_{(r)} &= \sum_{i=t+1}^{n'} \eta_i f(i;n',\epsilon'_{(r)}), \\
   \gamma_{(r)} &= \sum_{i=t+1}^{n'} (1-\eta_i) f(i;n',\epsilon'_{(r)}),
   \end{align*}
where $F$ and $f$ denote the CDF and PMF of the binomial distribution, respectively, as defined in Section~\ref{sec2}. For \mbox{$i \in [t+1,n']$}, the coefficients $\eta_i \in (0,1)$ quantify the fraction of error patterns of Hamming weight $i$ that result in a miscorrection rather than a decoding failure. Exact closed-form expressions of $\eta_i$ for linear MDS codes are provided in~\cite{cheung1989more}.
\end{lemma}
The results in Lemma~\ref{lemma2} apply to MDS codes in general, including Reed-Solomon codes, for which BDD can be implemented efficiently with quadratic complexity in $n'$~\cite{berlekamp1984algebraic}. The probabilities in Lemma~\ref{lemma2} are key to analyzing the performance of the outer MDS code and evaluating the probability of successful data retrieval, as we discuss in the next section. 

\subsection{Outer MDS Code and Data Retrieval} \label{sec3b}
A decoding failure in the inner MDS code results in a sequence dropout, which the outer MDS code can detect and treat as an erasure. In contrast, a miscorrection in the inner code remains undetected and leads to incorrect sequence information, effectively causing substitution errors in the outer code. The outer $(N, K)$~MDS code can correct any combination of $e^{\text{era}}$ erasures and $e^{\text{sub}}$ substitutions, provided that \mbox{$e^{\text{era}} + 2e^{\text{sub}} \leq N - K$}, which is a standard property of MDS codes. Equivalently, the data can be retrieved successfully if the number of correctly decoded sequences exceeds the number of incorrectly decoded ones by at least~$K$. To formulate this condition mathematically, we introduce the following definition.

\begin{definition}[Inner-decoder indicator variables and their sum]
\label{def:SN}
Let $\boldsymbol{r}=(r_1,\ldots,r_N)$ denote the read profile of the $N$ encoded sequences $(\boldsymbol{x}^j)_{j=1}^N$, where $r_j$ is the number of reads of $\tilde{\boldsymbol{x}}_j$. For each $j\in[N]$, we define a random variable \mbox{$\rv{Z}_j(r_j)\in\{-1,0,1\}$} indicating the inner-decoding outcome (see Definition~\ref{def:abg}), with
\begin{align*}
    \Pr(\rv{Z}_j(r_j)=1)&=\alpha_{(r_j)} \\
    \Pr(\rv{Z}_j(r_j)=-1)&=\beta_{(r_j)}, \\
    \Pr(\rv{Z}_j(r_j)=0)&=\gamma_{(r_j)}.
\end{align*}
We further define the sum
\[
\rv{S}_N(\boldsymbol{r}) \triangleq \rv{Z}_1(r_1) + \rv{Z}_2(r_2) + \ldots + \rv{Z}_N(r_N),
\]
which aggregates the inner-decoder outcomes across all $N$ sequences. In the special case of a uniform read profile, where $r_j=r$ for all $j\in[N]$, we write $\rv{S}_N(r)$ instead of $\rv{S}_N(\boldsymbol{r})$.
\end{definition}

The decoding of the outer $(N, K)$ MDS code is successful, and the data is thus successfully retrieved, if \mbox{$\rv{S}_N{(\boldsymbol{r})} \geq K$}. This condition ensures that the total number of errors and erasures resulting from the inner code are within the error correction capability of the outer MDS code, forming the basis of our result in Theorem~\ref{thm1}.

\begin{theorem} \label{thm1}
For a given read profile $\boldsymbol{r}=(r_1,\ldots,r_N)$, the probability of successful retrieval $P_{\text{\normalfont succ}\mid\boldsymbol{r}}$ satisfies
      \begin{equation*}
       P_{\text{\normalfont succ}\mid\boldsymbol{r}}\geq \Pr\left(\rv{S}_N{(\boldsymbol{r})}\geq K \right)= \hspace{-0.8cm} \sum_{\substack{\boldsymbol{z} \in \{-1,0,1\}^N \\ K\leq z_1+\ldots+z_N \leq N }} \hspace{-0.1cm} \prod_{j=1}^N \Pr\left(\rv{Z}_j{(r_j)} = z_j \right),
   \end{equation*}
  which can alternatively be expressed as in~\eqref{eqP0}, where $N$ and $K$ are the parameters of the outer $(N,K)$ MDS code. 
\end{theorem}

In the special case where all encoded sequences are read an equal number of times, the result in Theorem~\ref{thm1} simplifies to the expressions given in Corollary~\ref{corr4}. While this scenario is idealistic and far from typical in practice, Corollary~\ref{corr4} will be useful for deriving subsequent results.

\begin{corollary} \label{corr4}
Consider a uniform read profile \mbox{$\boldsymbol{r}=(r_1,\ldots,r_N)$}, with $r_j=r$ for all $j\in[N]$. The PMF of the random variable \mbox{$\rv{S}_N{(r)}\in [-N, N]$} is given by 
\begin{equation*}
 \Pr\left(\rv{S}= s \right) = \hspace{-0.5cm} \sum_{i = \max \{-s , 0\}}^{\lfloor (N-s) / 2 \rfloor}  \hspace{-0.2cm}  \binom{N}{i+s,i,N-2i-s} \alpha_{(r)}^{i+s} \beta_{(r)}^{i} \gamma_{(r)}^{N-2i-s},
\end{equation*}
and consequently 
\begin{equation*}
P_{\text{\normalfont succ}\mid r} \geq \sum_{s=K}^N \Pr\left(\rv{S}_N{(r)}= s \right).
\end{equation*}
\end{corollary}

\begin{remark} \label{rem:score_condition}
The condition \( \rv{S}_N(\boldsymbol{r}) \ge K \) in Theorem~\ref{thm1} is sufficient but not necessary, which is why it gives a lower bound on \( P_{\text{\normalfont succ} \mid \boldsymbol{r}} \). Recall that the outer $(N,K)$ MDS code operates column-wise over $\mathbb{F}_{2^M}$ with $k/M$ columns (see Fig.~\ref{fig:scheme}). Thus, a necessary and sufficient condition for successful retrieval is
\begin{equation}
e_l^{\text{era}} + 2e_l^{\text{sub}} \le N - K, \qquad \forall\, l \in [k/M],
\label{eq:necessary_and_sufficient}
\end{equation}
where $e_l^{\text{era}}$ and $e_l^{\text{sub}}$ are the numbers of erasures and substitutions in column $l$. One can easily show that \( \rv{S}_N(\boldsymbol{r}) \ge K \) implies~\eqref{eq:necessary_and_sufficient}, but the converse is not generally true. For instance, a single substitution in any row reduces \( \rv{S}_N \) by one, even if the remaining columns are error-free. Thus,~\eqref{eq:necessary_and_sufficient} may still hold when \( \rv{S}_N(\boldsymbol{r}) < K \),  due to a favorable dispersion of substitution errors across columns. In the special case $M = k$ (single column), the two conditions are equivalent, and the bound in Theorem~\ref{thm1} holds with equality, i.e., $P_{\text{\normalfont succ} \mid \boldsymbol{r}} = \Pr\left(\rv{S}_N(\boldsymbol{r}) \ge K\right)$. In general, the condition \( \rv{S}_N(\boldsymbol{r}) \ge K \) offers analytical simplicity, and Theorem~\ref{thm1} holds for any value of $M$, provided that $2^M\geq N$.
\end{remark}

The importance of Theorem~\ref{thm1} is that it enables the analytical evaluation of the probability of successful data retrieval for a given read profile. Despite the exponential number of subsets involved in the expression given in~\eqref{eqP0}, this probability can be efficiently computed by deriving a recurrence relation for the PMF of $\rv{S}_N{(\boldsymbol{r})}$, as we show next. To this end, we introduce partial sums of inner-decoding outcomes in Definition~\ref{def:Sj}.

\begin{definition}[Partial sums of inner-decoding outcomes]
\label{def:Sj}
For $j\in[N]$ and $\boldsymbol{r}_{[1,j]}=(r_1,\ldots,r_j)$, we define
\[
\rv{S}_j(\boldsymbol{r}_{[1,j]}) \triangleq \rv{Z}_1(r_1) + \rv{Z}_2(r_2) + \ldots + \rv{Z}_j(r_j).
\]
For brevity, we write $\rv{S}_j$ as shorthand for $\rv{S}_j{(\boldsymbol{r}_{[1,j]})}$. 
\end{definition}

\noindent The following recurrence relation holds for \mbox{$j\in [N]$}:
\begin{multline} \label{eq4}
\Pr\left(\rv{S}_j= s \right) =  \alpha_{(r_j)} \Pr\left(\rv{S}_{j-1}= s-1 \right)  \\ + \beta_{(r_j)} \Pr\left(\rv{S}_{j-1}= s+1 \right) + \gamma_{(r_j)} \Pr\left(\rv{S}_{j-1}= s \right), 
\end{multline}
with $\Pr\left(\rv{S}_0= s \right) = \mathds{1}_{\{s = 0\}}$. The PMF of \mbox{$\rv{S}_N{(\boldsymbol{r})} \in [-N, N]$}, given by $\Pr\left(\rv{S}_N= s \right)$, can thus be computed in $\mathcal{O}(N^2)$ time using the recurrence relation in~\eqref{eq4}. Once this PMF is determined, the bound in Theorem~\ref{thm1} is obtained by summing its upper tail, namely $\Pr\!\left(\rv{S}_N(\boldsymbol{r}) \ge K\right) = \sum_{s=K}^{N} \Pr(\rv{S}_N = s)$.

Alternatively, since $\rv{Z}_{1}{(r_1)},\ldots,\rv{Z}_{N}{(r_N)}$, are independent random variables for a given~$\boldsymbol{r}$, we can also approximate $\Pr\left(\rv{S}_N{(\boldsymbol{r})}\geq K \right)$ in terms of the CDF of the standard Gaussian distribution.  This follows from generalized versions of the central limit theorem (CLT), which apply to sums of independent but non-identically distributed random variables under certain conditions. For large~$N$, this approximation allows evaluating $\Pr\left(\rv{S}_N{(\boldsymbol{r})}\geq K \right)$ in $\mathcal{O}(N)$ time. The resulting CLT-based approximation is formalized in Corollary~\ref{corr5}.

\begin{corollary} \label{corr5}
The mean and variance of $\rv{S}_N{(\boldsymbol{r})}$ are 
  \begin{align*}
      \mu_{(\boldsymbol{r})} &= \sum_{j=1}^N \mathbb{E}[\rv{Z}_j(r_j)] =  \sum_{j=1}^N \alpha_{(r_j)}-\beta_{(r_j)}, \\
      \sigma_{(\boldsymbol{r})}^2 &= \sum_{j=1}^N \mathbb{V} [\rv{Z}_j(r_j)] = \sum_{j=1}^N \alpha_{(r_j)}+\beta_{(r_j)}-\left(\alpha_{(r_j)}-\beta_{(r_j)}\right)^2.
  \end{align*}
If $\|\boldsymbol{r} \|_0\geq N^{\theta}$ for some constant $\theta\in (0,1]$, then for large $N$, we have $$\Pr\left(\rv{S}_N{(\boldsymbol{r})}\geq K \right) \approx \Phi\left(\frac{\mu_{(\boldsymbol{r})}-K+0.5}{\sigma_{(\boldsymbol{r})}}\right),$$
where $\Phi$ is the CDF of a standard Gaussian distribution. The approximation error is bounded by
$$\left \lvert  \Pr\left(\rv{S}_N{(\boldsymbol{r})}\geq K \right) -  \Phi\left(\frac{\mu_{(\boldsymbol{r})}-K+0.5}{\sigma_{(\boldsymbol{r})}}\right) \right \rvert \leq  \frac{C \zeta_{(\boldsymbol{r})}}{ \sqrt{N} \sigma_{(\boldsymbol{r})}^{3/2}} ,$$
where $C\leq 0.5606$ is a constant, and $\zeta_{(\boldsymbol{r})}$ denotes the sum of the third absolute central moments of $\rv{Z}_{1}{(r_1)},\ldots,\rv{Z}_{N}{(r_N)}$.
\end{corollary}


\subsection{Data Retrieval as a Function of the Total Number of Reads} \label{sec3c}

In the previous sections, we focused on the randomness introduced by substitution errors and analyzed the probability of successful data retrieval $P_{\text{\normalfont succ}\mid \boldsymbol{r}}$ for a given read profile~$\boldsymbol{r}$. Another important source of randomness arises from the sequencing process itself, which determines the statistical behavior of $\boldsymbol{r}$.  We now extend our analysis to incorporate the randomness in~$\boldsymbol{r}$ and study its impact on the reliability of data retrieval. Under the multinomial model (Section~\ref{channel}), where a total of $R_{\text{\normalfont all}}$ samples (reads) are drawn from a probability distribution $\boldsymbol{p}$, the law of total probability gives
\begin{equation} \label{eqpR}
    P_{\text{succ}\mid\boldsymbol{p},R_{\text{\normalfont all}}} =\sum_{\substack{\boldsymbol{r} \in \mathbb{N}^N \\  r_1+\ldots+r_N=R_{\text{\normalfont all}}}} P_{\text{succ} \mid \boldsymbol{r}} \binom{R_{\text{\normalfont all}}}{r_1,\ldots,r_N} \prod_{j=1}^N p_j^{r_j}.
\end{equation}

Evaluating $P_{\text{succ}\mid\boldsymbol{p},R_{\text{\normalfont all}}}$ from~\eqref{eqpR} is computationally challenging, even for moderate values of $N$ and $R_{\text{\normalfont all}}$, due to the combinatorial explosion in the number of summation terms. A practical alternative is to evaluate it numerically by computing the empirical mean of \( P_{\text{succ} \mid \boldsymbol{r}} \) over a large number of read profiles \( \boldsymbol{r} \) drawn from the multinomial distribution. Analytically, various simplifications and bounding techniques have been proposed in the literature that facilitate the analysis\cite{10750859,preuss2024sequencing}. These simplifications rely on assumptions about $\boldsymbol{p}$ and $P_{\text{succ} \mid \boldsymbol{r}}$, such as: \begin{enumerate*}[label={\textit{(\roman*)}}] \item assuming $\boldsymbol{p}$ is uniform (i.e., $p_j=1/N$ for all~$j$); and/or \item treating $P_{\text{succ} \mid \boldsymbol{r}}$ as binary (e.g., decoding a sequence succeeds with probability one if its number of reads exceeds a predefined threshold, and with probability zero otherwise) \end{enumerate*}. These assumptions reduce the problem to well-known formulations like the {\em coupon collector} or {\em double dixie cup} problems. Bounds, such as $P_{\text{succ} \mid \boldsymbol{r}} \geq P_{\text{succ} \mid r_{\text{min}}}$, where \mbox{$r_{\text{min}} = \min\{r_1,\ldots,r_N\}$}, can also help simplify the analysis.

While the aforementioned assumptions facilitate analysis and provide valuable theoretical insights, they generally do not reflect practical realities. Experimental studies, such as~\cite{gimpel2023digital}, highlight the prevalence of biases, whereby synthesis inefficiencies, amplification variability, and strand degradation can lead to substantial variations in physical coverage.  Consequently, $\boldsymbol{p}$ is typically non-uniform, resulting in some encoded sequences being overrepresented, underrepresented, or absent in the sequencer output. This non-uniformity can also render bounds like \mbox{$P_{\text{succ} \mid \boldsymbol{r}} \geq P_{\text{succ} \mid r_{\text{min}}}$} loose or trivial when $r_{\text{min}}$ is very small or zero, especially if $R_{\text{\normalfont all}} = \mathcal{O}(N)$. Furthermore, as discussed in Sections~\ref{sec3a} and~\ref{sec3b}, $P_{\text{succ} \mid \boldsymbol{r}}$ is not binary in practice due to the randomness of errors.

To explore a more general framework for reliability that accounts for randomness in both sequencing and error processes, while allowing for biases, we avoid restrictive assumptions on $\boldsymbol{p}$ and $P_{\text{succ} \mid \boldsymbol{r}}$. One of the interesting applications of this framework, discussed further in Section~\ref{opt}, is to determine the minimum number of reads required to ensure successful data retrieval with a probability exceeding a desired threshold.  In addressing such practical questions, it is useful to establish a computationally tractable lower bound on the probability of success. To this end, we next introduce a bound that overcomes the computational bottleneck of evaluating~\eqref{eqpR}, which we later use in Section~\ref{opt} to analyze various trade-offs and optimization~problems.

\begin{definition}[Read frequency]
\label{def:readfreq}
Let $(\rv{R}_1,\ldots,\rv{R}_N)$ be the random read profile associated with the $N$ encoded sequences. We define the read frequency as the random vector 
\mbox{$(\rv{H}_0,\rv{H}_1,\ldots,\rv{H}_{R_{\text{\normalfont all}}})$}, where
\[
\rv{H}_i \triangleq \sum_{j=1}^{N} \mathds{1}_{\{\rv{R}_j = i\}},
\quad i \in \{0,1,\ldots,R_{\text{\normalfont all}}\},
\]
denotes the number of encoded sequences that are read exactly~$i$~times.
We further define the cumulative read frequency
\[
\tilde{\rv{H}}_i \triangleq \sum_{l=0}^{i} \rv{H}_l,
\]
which corresponds to the number of encoded sequences read at most $i$ times, with $\tilde{\rv{H}}_{R_{\text{\normalfont all}}}=N$.
\end{definition}

Our two-step approach for deriving the lower bound is as follows:
\begin{enumerate}[leftmargin=*]
\item We establish a lower bound of the form \mbox{$P_{\text{succ} \mid  \boldsymbol{r}}\geq P_{\text{succ} \mid  h_0,\tilde{h}_{r'}}$}, which depends on $\boldsymbol{r}$ only through~$h_0$ (the number of sequence dropouts) and $\tilde{h}_{r'}$ (the number of encoded sequences that are read at most $r'$ times, for some fixed~$r'$).
\item We derive a recurrence relation to compute the joint PMF of $\rv{H}_0$ and $\tilde{\rv{H}}_{r'}$ under the Poisson model (Section~\ref{channel}). Then, using the bound \mbox{$P_{\text{succ} \mid  \boldsymbol{r}}\geq P_{\text{succ} \mid  h_0,\tilde{h}_{r'}}$}, we apply the law of total probability to marginalize over the pair $(\rv{H}_0, \tilde{\rv{H}}_{r'})$.
\end{enumerate}

The lower bound derived in the first step is given in Lemma~\ref{LB1}. The importance of Lemma~\ref{LB1} lies in reducing the analysis to two random variables, $\rv{H}_0$ and $\tilde{\rv{H}}_{r'}$, instead of the entire random vector $(\rv{R}_1,\ldots,\rv{R}_N)$, thereby simplifying the application of the law of total probability in the second step.

\begin{lemma} \label{LB1}
Consider a given read profile \mbox{$\boldsymbol{r}=(r_1,\ldots,r_N)$}, and let $\boldsymbol{h}=(h_0,h_1,\ldots,h_{R_{\text{\normalfont all}}})$ be its corresponding frequency vector, where $R_{\text{\normalfont all}}=r_1+\ldots+r_N$. For any $r' \in [R_{\text{\normalfont all}}]$ and any subset \mbox{$\mathcal{S}\subseteq [-(\tilde{h}_{r'}-h_0),\tilde{h}_{r'}-h_0]$}, the following lower bound holds
\begin{equation} \label{eqLBh}
P_{\text{\normalfont succ} \mid  \boldsymbol{r}} \geq \textstyle \sum_{s\in \mathcal{S}}  \Pr (\rv{S}'_{\text{\normalfont high}} \geq K-s \mid \rv{S}'_{\text{\normalfont low}}=s ) \Pr(\rv{S}'_{\text{\normalfont low}} = s),
\end{equation}
where $\rv{S}'_{\text{\normalfont low}}$ and $\rv{S}'_{\text{\normalfont high}}$ are the random variables defined by
$$
\rv{S}'_{\text{\normalfont low}} 
\;\triangleq\; 
\sum_{j=h_0+1}^{\tilde{h}_{r'}} \rv{Z}_{j}(1),
\quad
\rv{S}'_{\text{\normalfont high}}
\;\triangleq\;
\sum_{j=\tilde{h}_{r'}+1}^{N} \rv{Z}_{j}(r'+1). $$
The PMFs of $\rv{S}'_{\text{\normalfont low}}$ and  $\rv{S}'_{\text{\normalfont high}}$ follow from Corollary~\ref{corr4} by substituting $(N,r)$ with $(\tilde{h}_{r'}-h_0,1)$ and $(N-\tilde{h}_{r'},r'+1)$, respectively.
\end{lemma}

\begin{figure*}[h!]
\centering
\begin{subfigure}[b]{0.33\textwidth}
\centering
\resizebox{\textwidth}{!}{%
\begin{tikzpicture}[scale=1]
\begin{axis}[
    legend cell align={left},
    tick scale binop=\times,
    xlabel={Sampling Probability},
    ylabel={Relative Frequencies},
    y label style={at={(axis description cs:0.06,.5)}, anchor=south},
    x label style={at={(axis description cs:0.5,0.025)}, anchor=north},
    grid=major,
    xmajorgrids=true,
    xminorgrids=false,
    ymajorgrids=true,
    yminorgrids=true,
    xtick={0, 1e-4, 2e-4, 3e-4, 4e-4},
    ytick={0, 0.02, 0.04, 0.06, 0.08, 0.1, 0.12, 0.14},
    y tick label style={/pgf/number format/fixed},
    yticklabels={0, 0.02, 0.04, 0.06, 0.08, 0.1, 0.12, 0.14},
    legend style={at={(0.97,0.97)},anchor=north east},
    legend cell align={left},
    xmin=0, xmax=4e-4,
    ymin=0, ymax=0.14,
]

\addplot+[
    ybar interval,
    mark=no,
    fill=col1fade,
    draw=col1,
    line width=1.2pt
] coordinates {
(0.00000, 0.00437)
(1.00e-05, 0.02233)
(2.00e-05, 0.04760)
(3.00e-05, 0.06175)
(4.00e-05, 0.07543)
(5.00e-05, 0.08446)
(6.00e-05, 0.08189)
(7.00e-05, 0.07923)
(8.00e-05, 0.08265)
(9.00e-05, 0.07192)
(1.00e-04, 0.06308)
(1.10e-04, 0.05273)
(1.20e-04, 0.04646)
(1.30e-04, 0.04199)
(1.40e-04, 0.03952)
(1.50e-04, 0.02860)
(1.60e-04, 0.02033)
(1.70e-04, 0.02014)
(1.80e-04, 0.01435)
(1.90e-04, 0.01245)
(2.00e-04, 0.00979)
(2.10e-04, 0.00817)
(2.20e-04, 0.00646)
(2.30e-04, 0.00447)
(2.40e-04, 0.00418)
(2.50e-04, 0.00390)
(2.60e-04, 0.00285)
(2.70e-04, 0.00200)
(2.80e-04, 0.00143)
(2.90e-04, 0.00095)
(3.00e-04, 0.00114)
(3.10e-04, 0.00057)
(3.20e-04, 0.00038)
(3.30e-04, 0.00048)
(3.40e-04, 0.00048)
(3.50e-04, 0.00029)
(3.60e-04, 0.00000)
(3.70e-04, 0.00048)
(3.80e-04, 0.00019)
(3.90e-04, 0.00010)
(4.0e-04, 0.00000)
};

\addplot+[
    ybar interval,
    mark=no,
    fill=none,
    draw=col2,
    line width=1.5pt
] coordinates {
(0.00000, 0.00000)
(1.00e-05, 0.00000)
(2.00e-05, 0.00238)
(3.00e-05, 0.01511)
(4.00e-05, 0.03772)
(5.00e-05, 0.07163)
(6.00e-05, 0.09548)
(7.00e-05, 0.11961)
(8.00e-05, 0.13661)
(9.00e-05, 0.12322)
(1.00e-04, 0.11011)
(1.10e-04, 0.08683)
(1.20e-04, 0.06793)
(1.30e-04, 0.04760)
(1.40e-04, 0.03164)
(1.50e-04, 0.01967)
(1.60e-04, 0.01397)
(1.70e-04, 0.00694)
(1.80e-04, 0.00580)
(1.90e-04, 0.00352)
(2.00e-04, 0.00190)
(2.10e-04, 0.00105)
(2.20e-04, 0.00095)
(2.30e-04, 0.00000)
(2.40e-04, 0.00019)
(2.50e-04, 0.00000)
(2.60e-04, 0.00010)
(2.70e-04, 0.00000)
(2.80e-04, 0.00010)
(2.90e-04, 0.00000)
(3.00e-04, 0.00000)
(3.10e-04, 0.00000)
(3.20e-04, 0.00000)
(3.30e-04, 0.00000)
(3.40e-04, 0.00000)
(3.50e-04, 0.00000)
(3.60e-04, 0.00000)
(3.70e-04, 0.00000)
(3.80e-04, 0.00000)
(3.90e-04, 0.00000)
(4.0e-04, 0.00000)
};

\addplot+[
    ybar interval,
    mark=no,
    fill=none,
    draw=none,
    postaction={
        pattern=north east lines,
        pattern color=col3
    },
    line width=0pt
] coordinates {
(2.00e-05, 0.00238)
(3.00e-05, 0.01511)
(4.00e-05, 0.03772)
(5.00e-05, 0.07163)
(6.00e-05, 0.08189)
(7.00e-05, 0.07923)
(8.00e-05, 0.08265)
(9.00e-05, 0.07192)
(1.00e-04, 0.06308)
(1.10e-04, 0.05273)
(1.20e-04, 0.04646)
(1.30e-04, 0.04199)
(1.40e-04, 0.03164)
(1.50e-04, 0.01967)
(1.60e-04, 0.01397)
(1.70e-04, 0.00694)
(1.80e-04, 0.00580)
(1.90e-04, 0.00352)
(2.00e-04, 0.00190)
(2.10e-04, 0.00105)
(2.20e-04, 0.00095)
(2.30e-04, 0.00000)
};

\legend{$\xi=3$, $\xi=9$}
\end{axis}
\end{tikzpicture}}
\caption{}
\label{fig2a}
\end{subfigure}
\hfill
\begin{subfigure}[b]{0.33\textwidth}
\centering
  \resizebox{\textwidth}{!}{%
  \begin{tikzpicture}[scale=1]
  \begin{axis}[
    legend cell align={left},
    tick scale binop=\times,
              y label style={at={(axis description cs:0.09,.5)}, anchor=south},
      x label style={at={(axis description cs:0.5,0.025)}, anchor=north},
    xlabel={Total Number of Reads $R_{\text{all}}$ },
    ylabel={Probability of Successful Retrieval},
    xticklabel={\pgfmathparse{\tick/1e-1}\pgfmathprintnumber{\pgfmathresult}},
    xtick scale label code/.code={$\times 10^4$}, 
    grid=both,
    ytick = {0, 0.1, 0.2, 0.3, 0.4, 0.5, 0.6, 0.7, 0.8, 0.9, 1},
    legend style={at={(0.75,0.8)}, font=\footnotesize},
    legend style={font=\footnotesize},
    ymin = 0,
    ymax = 1,
    xmax = 100000,
    xmin = 50000,
    minor x tick num=1,
    minor y tick num=1,
    xmajorgrids=true,
    xminorgrids=true,
    ymajorgrids=true,
    yminorgrids=true,
    major grid style={black!30}, 
    minor grid style={dotted,black!50}, 
]

\addplot+[color=blue, dashed, mark=triangle,mark size=1.5, mark options={solid, fill=none},line width=0.6pt, ] coordinates {
(80550, 2.41e-06)
(81050, 1.04e-05)
(81550, 4.06e-05)
(82050, 0.000142422)
(82550, 0.000451456)
(83050, 0.001297704)
(83550, 0.003392765)
(84050, 0.00809238)
(84550, 0.017665015)
(85050, 0.035407634)
(85550, 0.06539322)
(86050, 0.111693761)
(86550, 0.177140445)
(87050, 0.26198938)
(87550, 0.363070913)
(88050, 0.473925673)
(88550, 0.58604123)
(89050, 0.69079274)
(89550, 0.781360569)
(90050, 0.85394211)
(90550, 0.907945341)
(91050, 0.945308298)
(91550, 0.969383038)
(92050, 0.983851971)
(92550, 0.991974789)
(93050, 0.99624057)
(93550, 0.998339144)
(94050, 0.999307605)
(94550, 0.999727419)
(95050, 0.999898587)
(95550, 0.999964313)
(96050, 0.999988111)
(96550, 0.999996246)
(97050, 0.999998875)
(97550, 0.999999679)
(98050, 0.999999912)
(98550, 0.999999976)
(99050, 0.999999993)
(99550, 0.999999997)
(100050, 0.999999998)
(100550, 0.999999998)
};

\addplot+[color=red,mark=o,mark size=1.5, mark options={fill=none},line width=0.5pt] coordinates {
(80550, 1.8e-06)
(81050, 8.62e-06)
(81550, 4.04e-05)
(82050, 0.000124)
(82550, 0.00044)
(83050, 0.0013)
(83550, 0.004)
(84050, 0.011)
(84550, 0.0171)
(85050, 0.039685381)
(85550, 0.07060525)
(86050, 0.126927413)
(86550, 0.189447319)
(87050, 0.292049752)
(87550, 0.380081425)
(88050, 0.4931471)
(88550, 0.618936146)
(89050, 0.722646758)
(89550, 0.801257857)
(90050, 0.877376182)
(90550, 0.92119907)
(91050, 0.959467628)
(91550, 0.973804834)
(92050, 0.988066533)
(92550, 0.993647975)
(93050, 0.997170556)
(93550, 0.998554112)
(94050, 0.999627335)
(94550, 0.999864077)
(95050, 0.999904932)
(95550, 0.999979046)
(96050, 0.999996094)
(96550, 0.999998716)
(97050, 0.999999751)
(97550, 0.999999934)
(98050, 0.99999999)
(98550, 0.999999996)
(99050, 1)
(99550, 1)
(100050, 1)
(100550, 1)
};

\addplot+[color=orange, dashed, mark=triangle,mark size=1.5, mark options={solid, fill=none},line width=0.6pt] coordinates {
(52777, 3.62e-19)
(53277, 9.65e-17)
(53777, 1.54e-14)
(54277, 1.53e-12)
(54777, 9.85e-11)
(55277, 4.21e-09)
(55777, 1.22e-07)
(56277, 2.42e-06)
(56777, 3.35e-05)
(57277, 0.000327624)
(57777, 0.002292183)
(58277, 0.011664794)
(58777, 0.043918722)
(59277, 0.124751497)
(59777, 0.273644205)
(60277, 0.477015421)
(60777, 0.684767737)
(61277, 0.844806229)
(61777, 0.938516879)
(62277, 0.980550286)
(62777, 0.99510142)
(63277, 0.999017564)
(63777, 0.999842761)
(64277, 0.999979846)
(64777, 0.999997921)
(65277, 0.999999825)
(65777, 0.999999987)
(66277, 0.999999998)
(66777, 0.999999998)
(67277, 0.999999998)
(67777, 0.999999998)
(68277, 0.999999998)
(68777, 0.999999999)
(69277, 0.999999999)
(69777, 0.999999999)
(70277, 0.999999999)
(70777, 0.999999999)
(71277, 0.999999999)
(71777, 0.999999999)
(72277, 0.999999999)
(72777, 0.999999999)
};

\addplot+[color=black,mark=o,mark size=1.5, mark options={fill=none},line width=0.5pt] coordinates {
(52777, 8.96e-23)
(53277, 1.13e-19)
(53777, 7.32e-17)
(54277, 6.83e-14)
(54777, 1.57e-11)
(55277, 2.24e-09)
(55777, 5.38e-08)
(56277, 1.11e-06)
(56777, 1.95e-05)
(57277, 0.000288636)
(57777, 0.001851949)
(58277, 0.010223039)
(58777, 0.045203626)
(59277, 0.123201515)
(59777, 0.292397748)
(60277, 0.49700592)
(60777, 0.712936265)
(61277, 0.877615708)
(61777, 0.950359924)
(62277, 0.988133793)
(62777, 0.996707435)
(63277, 0.99939433)
(63777, 0.999956954)
(64277, 0.999990518)
(64777, 0.999999347)
(65277, 0.999999961)
(65777, 0.999999999)
(66277, 1)
(66777, 1)
(67277, 1)
(67777, 1)
(68277, 1)
(68777, 1)
(69277, 1)
(69777, 1)
(70277, 1)
(70777, 1)
(71277, 1)
(71777, 1)
(72277, 1)
(72777, 1)
};

\legend{}
\addlegendentry{$\xi=3$, Analytical~\eqref{eqLBr}}
\addlegendentry{$\xi=3$, Numerical~\eqref{eqpR}}
\addlegendentry{$\xi=9$, Analytical~\eqref{eqLBr}}
\addlegendentry{$\xi=9$, Numerical~\eqref{eqpR}}

\end{axis}
  \end{tikzpicture}
  }
  \caption{}
  \label{fig2b}
 \end{subfigure}\hfill
\begin{subfigure}[b]{0.33\textwidth}
\centering
\resizebox{\textwidth}{!}{%
  \begin{tikzpicture}[scale=1]
  \begin{semilogyaxis}[
    legend cell align={left},
    tick scale binop=\times,
              y label style={at={(axis description cs:0.055,.5)}, anchor=south},
      x label style={at={(axis description cs:0.5,0.025)}, anchor=north},
    xlabel={Total Number of Reads $R_{\text{all}}$ },
    ylabel={Probability of Retrieval Error},
    xticklabel={\pgfmathparse{\tick/1e-1}\pgfmathprintnumber{\pgfmathresult}},
    xtick scale label code/.code={$\times 10^4$}, 
    grid=both,
    ymin=1e-7,
    ymax=1,
    xmax=100000,
    xmin=50000,
    minor x tick num=1,
    legend style={at={(0.85,0.4)}, font=\footnotesize},
    xmajorgrids=true,
    xminorgrids=true,
    ymajorgrids=true,
    yminorgrids=true,
    major grid style={black!30}, 
    minor grid style={dotted,black!50}, 
]

\addplot+[color=blue, dashed, mark=triangle,mark size=1.5, mark options={solid, fill=none},line width=0.6pt, ] coordinates {
(80550, 0.99999759)
(81050, 0.9999896)
(81550, 0.9999594)
(82050, 0.999857578)
(82550, 0.999548544)
(83050, 0.998702296)
(83550, 0.996607235)
(84050, 0.99190762)
(84550, 0.982334985)
(85050, 0.964592366)
(85550, 0.93460678)
(86050, 0.888306239)
(86550, 0.822859555)
(87050, 0.73801062)
(87550, 0.636929087)
(88050, 0.526074327)
(88550, 0.41395877)
(89050, 0.30920726)
(89550, 0.218639431)
(90050, 0.14605789)
(90550, 0.092054659)
(91050, 0.054691702)
(91550, 0.030616962)
(92050, 0.016148029)
(92550, 0.008025211)
(93050, 0.00375943)
(93550, 0.001660856)
(94050, 0.000692395)
(94550, 0.000272581)
(95050, 0.000101413)
(95550, 3.5687e-05)
(96050, 1.1889e-05)
(96550, 3.754e-06)
(97050, 1.125e-06)
(97550, 3.21e-07)
(98050, 8.799999995e-08)
(98550, 2.399999999e-08)
(99050, 7.000000024e-09)
(99550, 3.000000026e-09)
(100050, 2.000000054e-09)
(100550, 2.000000054e-09)
};

\addplot+[color=red,mark=o,mark size=1.5, mark options={fill=none},line width=0.5pt] coordinates {
(80550, 0.9999982)
(81050, 0.99999138)
(81550, 0.9999596)
(82050, 0.999876)
(82550, 0.99956)
(83050, 0.9987)
(83550, 0.996)
(84050, 0.989)
(84550, 0.9829)
(85050, 0.960314619)
(85550, 0.92939475)
(86050, 0.873072587)
(86550, 0.810552681)
(87050, 0.707950248)
(87550, 0.619918575)
(88050, 0.5068529)
(88550, 0.381063854)
(89050, 0.277353242)
(89550, 0.198742143)
(90050, 0.122623818)
(90550, 0.07880093)
(91050, 0.040532372)
(91550, 0.026195166)
(92050, 0.011933467)
(92550, 0.006352025)
(93050, 0.002829444)
(93550, 0.001445888)
(94050, 0.000372665)
(94550, 0.000135923)
(95050, 9.5068e-05)
(95550, 2.0954e-05)
(96050, 3.906e-06)
(96550, 1.284e-06)
(97050, 2.49e-07)
(97550, 6.600000002e-08)
(98050, 1.000000005e-08)
(98550, 3.999999998e-09)
(99050, 0)
(99550, 0)
(100050, 0)
(100550, 0)
};

\addplot+[color=orange, dashed, mark=triangle,mark size=1.5, mark options={solid, fill=none},line width=0.6pt] coordinates {
(52777, 1)
(53277, 1)
(53777, 1)
(54277, 1)
(54777, 0.9999999999)
(55277, 0.9999999958)
(55777, 0.999999878)
(56277, 0.99999758)
(56777, 0.9999665)
(57277, 0.999672376)
(57777, 0.997707817)
(58277, 0.988335206)
(58777, 0.956081278)
(59277, 0.875248503)
(59777, 0.726355795)
(60277, 0.522984579)
(60777, 0.315232263)
(61277, 0.155193771)
(61777, 0.061483121)
(62277, 0.019449714)
(62777, 0.00489858)
(63277, 0.000982436)
(63777, 0.000157239)
(64277, 2.0154e-05)
(64777, 2.079e-06)
(65277, 1.75e-07)
(65777, 1.299999997e-08)
(66277, 2.000000054e-09)
(66777, 2.000000054e-09)
(67277, 2.000000054e-09)
(67777, 2.000000054e-09)
(68277, 2.000000054e-09)
(68777, 9.999999717e-10)
(69277, 9.999999717e-10)
(69777, 9.999999717e-10)
(70277, 9.999999717e-10)
(70777, 9.999999717e-10)
(71277, 9.999999717e-10)
(71777, 9.999999717e-10)
(72277, 9.999999717e-10)
(72777, 9.999999717e-10)
};

\addplot+[color=black,mark=o,mark size=1.5, mark options={fill=none},line width=0.5pt] coordinates {
(52777, 1)
(53277, 1)
(53777, 1)
(54277, 1)
(54777, 1)
(55277, 0.9999999978)
(55777, 0.9999999462)
(56277, 0.99999889)
(56777, 0.9999805)
(57277, 0.999711364)
(57777, 0.998148051)
(58277, 0.989776961)
(58777, 0.954796374)
(59277, 0.876798485)
(59777, 0.707602252)
(60277, 0.50299408)
(60777, 0.287063735)
(61277, 0.122384292)
(61777, 0.049640076)
(62277, 0.011866207)
(62777, 0.003292565)
(63277, 0.00060567)
(63777, 4.3046e-05)
(64277, 9.482e-06)
(64777, 6.53e-07)
(65277, 3.900000001e-08)
(65777, 9.999999717e-10)
(66277, 0)
(66777, 0)
(67277, 0)
(67777, 0)
(68277, 0)
(68777, 0)
(69277, 0)
(69777, 0)
(70277, 0)
(70777, 0)
(71277, 0)
(71777, 0)
(72277, 0)
(72777, 0)
};

\legend{}
\addlegendentry{$\xi=3$, Analytical~\eqref{eqLBr}}
\addlegendentry{$\xi=3$, Numerical~\eqref{eqpR}}
\addlegendentry{$\xi=9$, Analytical~\eqref{eqLBr}}
\addlegendentry{$\xi=9$, Numerical~\eqref{eqpR}}

\end{semilogyaxis}
  \end{tikzpicture}
  }
  \caption{}
  \label{fig2c}
  \end{subfigure}
\caption{Retrieval performance under Dirichlet-distributed sampling probabilities as a function of the total number of reads $R_{\text{all}}$. (a)~Normalized histograms of two probability vectors drawn from a symmetric $N$-dimensional Dirichlet distribution $\text{Dir}_N(\xi)$, with $N=10526$ and $\xi=3,9$. (b)~Probability of successful retrieval and (c)~probability of retrieval error, evaluated for the sampling probability vectors in~(a) using a numerical approximation of~\eqref{eqpR} and the analytical bound from~\eqref{eqLBr}.}
\label{fig2}
\vspace{-0.4cm}
\end{figure*}

\begin{remark}
The bound in \eqref{eqLBh} is valid for any \(r' \in [R_{\text{all}}]\), so one may choose \(r'\) to maximize it. Given a read profile \(\boldsymbol{r}\), the choice of $r'$ divides the \(N\) encoded sequences into two disjoint classes: those with read count at least one and at most \(r'\); and those with at least \(r'+1\) reads. For the sake of the bounding argument, we assume all sequences in the first class as being read exactly once (defining \(\rv{S}'_{\text{low}}\)), and all sequences in the second class as being read exactly \(r' + 1\) times (defining \(\rv{S}'_{\text{high}}\)). A good way to select \(r'\) is to look for a large jump in \(\alpha_{(r'+1)} - \alpha_{(r')}\), where \(\alpha_{(r)} = \Pr(\rv{Z}(r) = +1)\). Since \(\alpha_{(r)}\) typically follows a sigmoidal pattern, setting \(r'\) near the threshold where \(\alpha_{(r)}\) increases sharply often yields a tight bound; this is illustrated through an example in Section~\ref{example}.
\end{remark}


To compute the joint PMF of $(\rv{H}_0,\tilde{\rv{H}}_{r'})$, we introduce partial read-frequency quantities in Definition~\ref{def:Hij}.

\begin{definition}[Partial read frequencies]
\label{def:Hij}
For $i\in\{0,1,\ldots,R_{\text{\normalfont all}}\}$ and $j\in[N]$, we define
\[
\rv{H}_{i,j} \triangleq \sum_{\ell=1}^{j} \mathds{1}_{\{\rv{R}_\ell = i\}}, \qquad
\tilde{\rv{H}}_{i,j} \triangleq \sum_{l=0}^{i} \rv{H}_{l,j},
\]
where $\rv{H}_{i,j}$ denotes the number of sequences among the first $j$ that are read exactly $i$ times, and $\tilde{\rv{H}}_{i,j}$ denotes the number that are read at most $i$ times.
\end{definition}
Let $g^{(j)}(h_0,\tilde{h}_{r'})\triangleq \Pr(\rv{H}_{0,j}=h_{0}, \tilde{\rv{H}}_{r',j} = \tilde{h}_{r'})$, for \mbox{$j\in[N]$}. Under the Poisson model described in Section~\ref{channel}, where \mbox{$\rv{R}_j \sim \text{Pois}(\lambda p_j)$} for $j\in [N]$, the following recurrence holds:
\begin{multline}\label{eqRecc}
g^{(j)}(h_0, \tilde{h}_{r'}) = (1 - q_0^{(j)} - q_{r'}^{(j)}) \, g^{(j-1)}(h_0,\tilde{h}_{r'})\\
+ q_{r'}^{(j)} \, g^{(j-1)}(h_0,\tilde{h}_{r'} - 1) + q_0^{(j)} \, g^{(j-1)}\bigl(h_0 - 1,\tilde{h}_{r'} - 1\bigr),
\end{multline}
where $$q_0^{(j)} \triangleq e^{-\lambda p_j}, \quad q_{r'}^{(j)} \triangleq \sum_{l=1}^{r'} \frac{(\lambda p_j)^l e^{-\lambda p_j}}{l!}.$$ 
The initial condition is $g^{(0)}(h_0,\tilde{h}_{r'})=1$ if $(h_{0},\tilde{h}_{r'})=(0,0)$, and zero otherwise. For $j=N$, we recover the joint PMF of $(\rv{H}_0, \tilde{\rv{H}}_{r'})$ from~\eqref{eqRecc}, which together with~\eqref{eqLBh}, leads to Theorem~\ref{thm2}.

\begin{theorem} \label{thm2}
Under the Poisson model described in Section~\ref{channel}, for any subsets
\(\mathcal{S}_0 \subseteq \mathcal{S}' \subseteq \mathbb{N}\), it holds that 
\begin{equation} \label{eqLBr}
P_{\text{\normalfont succ}\mid\boldsymbol{p},\lambda} \geq \textstyle \sum_{h_0\in \mathcal{S}_0} \sum_{h'\in \mathcal{S}'} P_{\text{\normalfont succ} \mid h_0,h'} \Pr(\rv{H}_0=h_0, \tilde{\rv{H}}_{r'}=h'),
\end{equation}
where $P_{\text{\normalfont succ} \mid h_0,h'}$ and $\Pr(h_0,h')$ follow from \eqref{eqLBh} and \eqref{eqRecc}, respectively.
\end{theorem}

The lower bound in Theorem~\ref{thm2} is derived under the Poisson model, which simplifies the analysis of the joint PMF of $(\rv{H}_0, \tilde{\rv{H}}_{r'})$ compared to the multinomial model. As explained in Section~\ref{channel}, these two models are related by Poissonization, and substituting $\lambda$ by $R_{\text{all}}$ in~\eqref{eqLBr} approximates the behavior of the multinomial model when $R_{\text{\normalfont all}}$ is large and the sampling probabilities are small (a regime typical for large DNA pools). In the next section, we consider an example where we evaluate both the analytical bound in~\eqref{eqLBr} and a numerical evaluation of~\eqref{eqpR} based on Theorem~\ref{thm1}.

\begin{figure*}[h!]
\centering
\begin{subfigure}[b]{0.33\textwidth}
\centering
\resizebox{\textwidth}{!}{%
  \begin{tikzpicture}[scale=1]
  \begin{axis}[
    legend cell align={left},
          y label style={at={(axis description cs:0.09,.5)}, anchor=south},
      x label style={at={(axis description cs:0.5,0.025)}, anchor=north},
    tick scale binop=\times,
          grid=major, 
      minor tick num=1,
    xlabel={Outer MDS Code Rate $\rho_{\text{out}}$},
    ylabel={Minimum Inf. Read Depth $R^{\star}_{\text{all}}/K$},
    grid=both,
    ytick={2,4,6,8,10,12,14,16,18,20,22,24,26},
    xtick={0.1,0.2,0.3,0.4,0.5,0.6,0.7,0.8,0.9,1},
    legend pos=north east,
    legend style={font=\small, xshift=-10pt},
    ymin = 0,
    ymax = 26,
    xmax = 1,
    xmin = 0.1,
    xmajorgrids=true,
    xminorgrids=true,
    ymajorgrids=true,
    yminorgrids=true,
    major grid style={black!30}, 
    minor grid style={dotted,black!50}, 
]

\addplot+[color=blue, mark=triangle,mark size=1.5, mark options={solid, fill=none},line width=0.6pt, ] coordinates {
(0.1, 25.211)
(0.15, 18.3742)
(0.2, 14.9859)
(0.25, 13.0258)
(0.3, 11.7267)
(0.35, 10.9184)
(0.4, 10.2934)
(0.45, 9.9446)
(0.5, 9.6911)
(0.55, 9.6003)
(0.6, 9.642)
(0.65, 9.799)
(0.7, 10.0898)
(0.75, 10.6273)
(0.8, 11.0728)
(0.85, 12.3921)
(0.9, 14.3737)
(0.95, 18.9634)
};

\addplot+[color=orange,mark=o,mark size=1.5, mark options={fill=none},line width=0.6pt] coordinates {
(0.1, 3.9096)
(0.15, 4.0553)
(0.2, 4.1571)
(0.25, 4.2398)
(0.3, 4.3172)
(0.35, 4.4059)
(0.4, 4.502)
(0.45, 4.6292)
(0.5, 4.7685)
(0.55, 4.9558)
(0.6, 5.1819)
(0.65, 5.4636)
(0.7, 5.821)
(0.75, 6.3147)
(0.8, 6.8449)
(0.85, 7.8347)
(0.9, 9.3763)
(0.95, 12.8419)
};

\addplot+[color=red, mark=square,mark size=1.5, mark options={solid, fill=none},line width=0.6pt] coordinates {
(0.1, 1.9797)
(0.15, 2.0692)
(0.2, 2.1662)
(0.25, 2.2737)
(0.3, 2.3845)
(0.35, 2.5126)
(0.4, 2.6452)
(0.45, 2.8034)
(0.5, 2.9691)
(0.55, 3.1692)
(0.6, 3.397)
(0.65, 3.668)
(0.7, 3.9974)
(0.75, 4.4278)
(0.8, 4.9165)
(0.85, 5.7456)
(0.9, 7.0273)
(0.95, 9.8552)
};

\legend{}
 \addlegendentry{$\rho_{\text{\normalfont in}}=1$}
\addlegendentry{$\rho_{\text{\normalfont in}} = 0.96$}
 \addlegendentry{$\rho_{\text{\normalfont in}}= 0.92$}

\end{axis}
  \end{tikzpicture}
  }
  \caption{$\epsilon=0.01$, $\xi=3$.}
  \label{fig3a}
 \end{subfigure}\hfill
\begin{subfigure}[b]{0.33\textwidth}
\centering
\resizebox{\textwidth}{!}{%
\begin{tikzpicture}[scale=1]
\begin{axis}[
    legend cell align={left},
    y label style={at={(axis description cs:0.09,.5)}, anchor=south},
    x label style={at={(axis description cs:0.5,0.025)}, anchor=north},
    tick scale binop=\times,
    xlabel={Outer MDS Code Rate $\rho_{\text{out}}$},
    ylabel={Minimum Inf. Read Depth $R^{\star}_{\text{all}}/K$},
    grid=both,
    ytick={2,4,6,8,10,12,14,16,18,20,22,24,26},
    xtick={0.1,0.2,0.3,0.4,0.5,0.6,0.7,0.8,0.9,1},
    legend pos=north east,
    legend style={font=\small, xshift=-30pt},
    ymin = 0,
    ymax = 26,
    xmax = 1,
    xmin = 0.1,
    xmajorgrids=true,
    xminorgrids=true,
    ymajorgrids=true,
    yminorgrids=true,
    minor tick num=1,
    major grid style={black!30},
    minor grid style={dotted,black!50},
]
\addplot+[color=red, mark=square, mark size=1.5, mark options={solid, fill=none}, line width=0.6pt] coordinates {
(0.10, 1.6792) (0.15, 1.7693) (0.20, 1.8665) (0.25, 1.9717) (0.30, 2.0881)
(0.35, 2.2149) (0.40, 2.3469) (0.45, 2.5027) (0.50, 2.6756) (0.55, 2.8679)
(0.60, 3.0856) (0.65, 3.3679) (0.70, 3.6953) (0.75, 4.1207) (0.80, 4.6572)
(0.85, 5.4315) (0.90, 6.7445) (0.95, 9.6778)
};
\addplot+[color=purple, mark=o, mark size=1.5, mark options={fill=none}, line width=0.6pt] coordinates {
(0.10, 10.8006) (0.15, 8.9050) (0.20, 7.9112) (0.25, 7.2978) (0.30, 6.9182)
(0.35, 6.6889) (0.40, 6.5303) (0.45, 6.4813) (0.50, 6.5011) (0.55, 6.5779)
(0.60, 6.7053) (0.65, 6.9683) (0.70, 7.3010) (0.75, 7.7970) (0.80, 8.4399)
(0.85, 9.4376) (0.90, 11.2278) (0.95, 15.3399)
};
\addplot+[color=teal, mark=triangle, mark size=1.5, mark options={solid, fill=none}, line width=0.6pt] coordinates {
(0.10, 16.1495) (0.15, 13.0923) (0.20, 11.4949) (0.25, 10.5024) (0.30, 9.8763)
(0.35, 9.4840) (0.40, 9.2010) (0.45, 9.0791) (0.50, 9.0592) (0.55, 9.1202)
(0.60, 9.2516) (0.65, 9.5666) (0.70, 9.9765) (0.75, 10.6016) (0.80, 11.4183)
(0.85, 12.7029) (0.90, 15.0214) (0.95, 20.3603)
};
\addplot+[color=custombrown, mark=diamond, mark size=1.5, mark options={solid, fill=none}, line width=0.6pt] coordinates {
(0.10, 23.6064) (0.15, 18.6721) (0.20, 16.1259) (0.25, 14.5412) (0.30, 13.5314)
(0.35, 12.8809) (0.40, 12.3960) (0.45, 12.1478) (0.50, 12.0450) (0.55, 12.0560)
(0.60, 12.1601) (0.65, 12.5064) (0.70, 12.9757) (0.75, 13.7172) (0.80, 14.6951)
(0.85, 16.2638) (0.90, 19.1150) (0.95, 25.7014)
};
\legend{}
\addlegendentry{$\epsilon = 0.01$}
\addlegendentry{$\epsilon = 0.05$}
\addlegendentry{$\epsilon = 0.10$}
\addlegendentry{$\epsilon = 0.15$}
\end{axis}
\end{tikzpicture}
  }
  \caption{$\rho_{\text{\normalfont in}}= 0.92$, $\xi=3$.}
  \label{fig3b}
 \end{subfigure}\hfill
\begin{subfigure}[b]{0.33\textwidth}
\centering
  \resizebox{\textwidth}{!}{%
  \begin{tikzpicture}[scale=1]
  \begin{axis}[
    legend cell align={left},
    tick scale binop=\times,
    xlabel={Dirichlet Bias Parameter $\xi$},
    ylabel={Minimum Inf. Read Depth $R^{\star}_{\text{all}}/K$},
          y label style={at={(axis description cs:0.09,.5)}, anchor=south},
      x label style={at={(axis description cs:0.5,0.025)}, anchor=north},
          grid=major, 
          minor x tick num=0,      
minor y tick num=1,      
xminorgrids=false,       
yminorgrids=true,        
    xtick={1,2,3,4,5,6,7,8,9,10},
    ytick={5,10,15,20,25,30,35,40,45,50,55,60},
    legend pos=north east,
    legend style={font=\small},
    ymin = 0,
    xmin=1,
    xmax = 10,
    ymax=60,
    xmajorgrids=true,
    xminorgrids=true,
    ymajorgrids=true,
    yminorgrids=true,
    major grid style={black!30}, 
    minor grid style={dotted,black!50}, 
]

\addplot+[color=blue, mark=triangle,mark size=1.5, mark options={solid, fill=none},line width=0.6pt, ] coordinates {
(1, 56.0934)
(2, 19.6147)
(3, 14.3657)
(4, 12.3249)
(5, 11.0913)
(6, 10.4514)
(7, 9.9607)
(8, 9.6204)
(9, 9.3666)
(10, 9.1652)
};

\addplot+[color=orange,mark=o,mark size=1.5, mark options={fill=none},line width=0.6pt] coordinates {
(1, 27.7549)
(2, 11.9123)
(3, 9.3621)
(4, 8.3578)
(5, 7.735)
(6, 7.4076)
(7, 7.1513)
(8, 6.9787)
(9, 6.8413)
(10, 6.7351)
};

\addplot+[color=red, mark=square,mark size=1.5, mark options={solid, fill=none},line width=0.6pt] coordinates {
(1, 18.9459)
(2, 8.6913)
(3, 7.016)
(4, 6.359)
(5, 5.9524)
(6, 5.7398)
(7, 5.5738)
(8, 5.462)
(9, 5.3742)
(10, 5.3058)
};
\legend{}
 \addlegendentry{$\rho_{\text{\normalfont in}}=1$}
\addlegendentry{$\rho_{\text{\normalfont in}} = 0.96$}
 \addlegendentry{$\rho_{\text{\normalfont in}}= 0.92$}

\end{axis}
  \end{tikzpicture}
  }
  \caption{$\epsilon=0.01$, $\rho_{\text{out}}=0.9$.}
  \label{fig3c}
  \end{subfigure}
\caption{Minimum information read depth \( R^\star_{\text{all}} / K \) required for reliable retrieval under the target threshold $\delta_{\text{th}}=10^{-6}$. (a) The outer MDS code rate \( \rho_{\text{out}} \) is varied for three inner code rates \( \rho_{\text{in}} = 1, 0.96, 0.92 \), with fixed channel error rate \( \epsilon = 0.01 \) and Dirichlet bias parameter \( \xi = 3 \). (b) The outer MDS code rate \( \rho_{\text{out}} \) is varied for four channel error rates \mbox{\( \epsilon = 0.01, 0.05, 0.10, 0.15 \)}, with fixed inner code rate \( \rho_{\text{in}} = 0.92 \) and Dirichlet bias parameter \( \xi = 3 \). (c) The Dirichlet bias parameter \( \xi \) is varied for three inner code rates \mbox{\( \rho_{\text{in}} = 1, 0.96, 0.92 \)}, with fixed channel error rate \( \epsilon = 0.01 \) and outer code rate \( \rho_{\text{out}} = 0.9 \).}
\label{fig3}
\vspace{-0.4cm}
\end{figure*}

\subsection{Example} \label{example}
Consider the following values of the parameters defined in Section~\ref{sec2}: an outer MDS code with $K=10000$, $N=10526$, and $\rho_{\text{out}}=0.95$;  an inner MDS code with $k=360$, $m=8$, $k'=45$, $n'=k'+2t=49$, and $\rho_{\text{in}}= 0.92$; and a QSC channel with $\epsilon=0.01$ (reported in~\cite{gimpel2023digital} as an upper limit for substitution error rates in certain practical setups). This configuration corresponds to storing a total of $N=10526$ DNA sequences, each of length $n'm/2=196$ NTs, with an information density $\Delta=2\rho_{\text{in}}\rho_{\text{out}}\approx 1.74$ bits/NT.

To account for biases, we draw the vector of sampling probabilities $\boldsymbol{p}=(p_1,\ldots,p_N)$ from a symmetric $N$-dimensional Dirichlet distribution with concentration parameter $\xi>0$, denoted~$\text{Dir}_N(\xi)$. Smaller values of $\xi$ (approaching zero) result in highly skewed sampling probabilities, while $\xi \to \infty$ approaches uniform sampling (i.e., $\boldsymbol{p}=(1/N,\ldots,1/N)$). Fig.~\ref{fig2a} shows the normalized histograms of two probability vectors $\boldsymbol{p}$ drawn from $\text{Dir}_N(\xi)$. The figure illustrates that for small $\xi$, a few encoded sequences attain relatively high sampling probabilities, producing a positive skew with a long right tail. Under the multinomial and Poisson models (see equations \eqref{eqM} and \eqref{eqP}), this skew propagates to the read distribution, whereby a small fraction of encoded sequences receive a significantly higher number of reads. Notably, such behavior aligns with experimental findings~\cite{erlich2017dna, chen2020quantifying, gimpel2023digital}.

We analyze the probability of successful retrieval as a function of the total number of reads $R_{\text{all}}$, using: \begin{enumerate*}[label={\textit{(\roman*)}}] \item the analytical lower bound in~\eqref{eqLBr} evaluated for $\lambda=R_{\text{all}}$, $r'=2$, and subsets that truncate the summations where the tail probabilities are negligible; and \item a numerical evaluation of~\eqref{eqpR} obtained by computing the empirical mean of $P_{\text{succ} \mid \boldsymbol{r}}$ based on  Theorem~\ref{thm1} over~$10^{3}$ read profiles \( \boldsymbol{r} \) drawn from the multinomial distribution. \end{enumerate*} We consider two values of $\xi$, $\xi=3$ and $\xi=9$, to explore how different levels of bias impact data retrieval. The results in Fig.~\ref{fig2} show that the probability of successful retrieval exhibits a sigmoidal behavior with respect to $R_{\text{all}}$, rising quickly to one near a threshold (Fig.~\ref{fig2b}), and causing the retrieval error probability to drop sharply (Fig.~\ref{fig2c}). The results also show good agreement between the analytical lower bound and the numerical evaluation, highlighting the effectiveness of the bound in Theorem~\ref{thm2}. Furthermore, when $\xi=3$, a significantly higher number of reads is required for reliable retrieval compared to $\xi=9$, due to the stronger skew in sampling probabilities (Fig.~\ref{fig2a}).

\section{Optimization and Trade-offs} \label{opt}
In this section, we apply the theoretical framework developed in Section~\ref{sec3} to analyze the trade‐offs between various design parameters that affect the reliability of data retrieval under MDS coding. Our goal is to identify optimal operating points that minimize sequencing or synthesis costs under reliability constraints. Specifically, we focus on two key optimization problems: \begin{enumerate*}[label={\textit{(\roman*)}}] \item minimizing the number of reads needed for retrieval (sequencing cost); and \item maximizing the information density achieved by the inner-outer code combination (synthesis cost);\end{enumerate*} both while ensuring a target success probability of at least $1-\delta_{\text{th}}$. We fix $\delta_{\text{th}}=10^{-6}$ throughout this section, and consider a similar configuration as in Section~\ref{example}, with $K=10000$, $k=360$, and $m=8$. To evaluate the optimal operating points, we use our analytical lower bound on the probability of success in~\eqref{eqLBr} (parametrized same as in Section~\ref{example}), and apply a grid search over the relevant parameters. 

\subsection{Minimum Number of Reads for Reliable Retrieval} \label{sec4a}
We measure sequencing cost through the total number of reads per information sequence \( R_{\text{all}}/K \), and therefore focus on the minimum number of reads \( R^{\star}_{\text{all}} \) required to achieve successful retrieval with probability exceeding the target threshold. We refer to \( R_{\text{all}}/K \) as the {\em information read depth}, which must be at least one to ensure a nonzero probability of successful retrieval.

Fig.~\ref{fig3a} illustrates the impact of the outer MDS code rate $\rho_{\text{out}}$ on $R^{\star}_{\text{all}}/K$ for different inner MDS code rates $\rho_{\text{in}}$, under a channel error rate $\epsilon = 0.01$. For $\rho_{\text{in}}=0.96$ and $0.92$, the sequencing cost decreases as more redundancy is introduced through the outer code, although with diminishing returns. This behavior is consistent with prior observations in the literature~\cite{10750859,abraham2024covering,10619151}. 

In contrast, for $\rho_{\text{in}}=1$, the results reveal an interesting nuance: adding redundancy in the outer code eventually becomes detrimental. The intuition here is that when $\rho_{\text{out}}$ is small, the total number of sequences $N$ is large, and the available reads are therefore distributed more thinly across the reference sequences. As a result, the consensus step becomes less effective, and a larger share of the error correction burden shifts to the inner code. When $\rho_{\text{in}}=1$, however, the inner code provides no protection beyond consensus, so the reliability of decoding individual sequences drops significantly and inner decoding becomes the bottleneck. Consequently, a larger overall number of reads is required to mitigate this effect and achieve reliable retrieval, leading to the behavior observed in Fig.~\ref{fig3a} for $\rho_{\text{in}}=1$.

This phenomenon is not limited to the case where $\rho_{\text{in}}=1$. Fig.~\ref{fig3b} shows that even when the inner code can correct errors, a similar trend appears as the channel error rate $\epsilon$ grows. For low error rates such as $\epsilon=0.01$, the inner code of rate $\rho_{\text{in}}=0.92$ provides sufficient protection for sequences receiving few reads, and increasing outer redundancy (i.e., decreasing $\rho_{\text{out}}$) therefore reduces the required information read depth. However, as the channel error rate increases while the inner code rate $\rho_{\text{in}}=0.92$ remains fixed, inner decoding becomes less reliable and the same tradeoff reappears.

Previous works on minimizing sequencing coverage depth~\cite{10750859,abraham2024covering,10619151} considered noiseless channel settings and concluded that increasing outer redundancy always reduces the required coverage, with the minimum attained as $\rho_{\text{out}} \to 0$. Our results reveal a more nuanced behavior in the presence of noise. Namely, the optimal operating point for minimizing coverage depth does not necessarily correspond to the lowest outer code rate. Instead, it depends on the interplay between channel error rate, inner code capability, and the read distribution. This insight opens new directions for future work on coding for minimizing sequencing costs under noisy conditions.

Finally, Fig.~\ref{fig3c} examines the impact of the Dirichlet bias parameter $\xi$, which governs the skewness of the sampling probabilities. As expected, stronger bias increases the required information read depth, since reads become more unevenly distributed across the reference sequences, which in turn makes reliable inner decoding more challenging.

\subsection{Optimal Redundancy Allocation}\label{sec4b}
We measure synthesis cost through the information density \mbox{$\Delta = 2\rho_{\text{in}}\rho_{\text{out}} = 2(k/n)(K/N)$}, and study the optimal redundancy allocation between the inner and outer codes that maximizes $\Delta$ under the same reliability constraint ($\delta_{\text{th}}=10^{-6}$).

Fig.~\ref{fig4a} shows the maximum achievable code rate \mbox{$\rho^{\star}=\rho_{\text{in}}^{\star}\rho_{\text{out}}^{\star}$} for $\epsilon=0.01$, as a function of the information read depth~$R_{\text{all}}/K$. The corresponding inner and outer code rates, $\rho_{\text{in}}^{\star}$ and $\rho_{\text{out}}^{\star}$, resulting from the optimal redundancy allocation are also reported. The results show that increasing the number of reads initially leads to a significant increase in $\rho^{\star}$, but this improvement gradually saturates due to diminishing returns. The maximum achievable information density follows as $\Delta^{\star}=2\rho^{\star}$. Our numerical results indicate that $\Delta^{\star}\approx 1.99$~bits/NT is achievable for a read depth of approximately $60$.

\begin{figure}[h!]
\centering
\begin{subfigure}[b]{0.49\columnwidth}
\centering
  \begin{tikzpicture}[scale=0.54]
  \begin{axis}[
      legend cell align={left},
      tick scale binop=\times,
      xlabel={Information Read Depth $R_{\text{all}}/K$},
      ylabel={Maximum Code Rate $\rho^{\star}$},
       y label style={at={(axis description cs:0.09,.5)}, anchor=south},
             x label style={at={(axis description cs:0.5,0.03)}, anchor=north},
      grid=major, 
      minor y tick num=1,
      minor x tick num=3,
      grid=both,
      ytick = {0.3,0.4,0.5,0.6,0.7,0.8,0.9,1},
      xtick = {1,3,5,7,9,11,13,15,17,19,21,23,25},
      legend pos=south east,
      legend style={font=\small},
      xticklabel style={/pgf/number format/fixed},
      xmax=25,
      xmin=1,
      ymax=1,
      xmajorgrids=true,
      xminorgrids=true,
      ymajorgrids=true,
      yminorgrids=true,
      major grid style={black!30},
      minor grid style={dotted,black!50},
    ]

    \addplot+[color=blue, mark=triangle,mark size=1.5, mark options={solid, fill=none},line width=0.6pt, ] coordinates {
(1.5, 0.789473684)
(2, 0.818181818)
(2.5, 0.818181818)
(3, 0.849056604)
(3.5, 0.849056604)
(4, 0.849056604)
(4.5, 0.882352941)
(5, 0.882352941)
(5.5, 0.882352941)
(6, 0.882352941)
(6.5, 0.882352941)
(7, 0.918367347)
(7.5, 0.918367347)
(8, 0.918367347)
(8.5, 0.918367347)
(9, 0.918367347)
(9.5, 0.918367347)
(10, 0.918367347)
(10.5, 0.957446809)
(11, 0.957446809)
(11.5, 0.957446809)
(12, 0.957446809)
(12.5, 0.957446809)
(13, 0.957446809)
(13.5, 0.957446809)
(14, 0.957446809)
(14.5, 0.957446809)
(15, 0.957446809)
(15.5, 0.957446809)
(16, 0.957446809)
(16.5, 1)
(17, 1)
(17.5, 1)
(18, 1)
(18.5, 1)
(19, 1)
(19.5, 1)
(20, 1)
(20.5, 1)
(21, 1)
(21.5, 1)
(22, 1)
(22.5, 1)
(23, 1)
(23.5, 1)
(24, 1)
(24.5, 1)
(25, 1)
};

\addplot+[color=red, mark=o,mark size=1.5, mark options={solid, fill=none},line width=0.6pt, ] coordinates {
(1.5, 0.415)
(2, 0.615)
(2.5, 0.735)
(3, 0.78)
(3.5, 0.83)
(4, 0.865)
(4.5, 0.86)
(5, 0.885)
(5.5, 0.9)
(6, 0.915)
(6.5, 0.93)
(7, 0.905)
(7.5, 0.915)
(8, 0.925)
(8.5, 0.935)
(9, 0.94)
(9.5, 0.945)
(10, 0.95)
(10.5, 0.92)
(11, 0.93)
(11.5, 0.935)
(12, 0.94)
(12.5, 0.945)
(13, 0.95)
(13.5, 0.955)
(14, 0.96)
(14.5, 0.96)
(15, 0.96)
(15.5, 0.965)
(16, 0.965)
(16.5, 0.93)
(17, 0.935)
(17.5, 0.935)
(18, 0.94)
(18.5, 0.945)
(19, 0.945)
(19.5, 0.95)
(20, 0.955)
(20.5, 0.955)
(21, 0.96)
(21.5, 0.96)
(22, 0.96)
(22.5, 0.96)
(23, 0.965)
(23.5, 0.965)
(24, 0.97)
(24.5, 0.97)
(25, 0.97)
};

\addplot+[color=magenta, mark=x,mark size=1.5, mark options={solid, fill=none},line width=0.6pt, ] coordinates {
(1.5, 0.327631579)
(2, 0.503181818)
(2.5, 0.601363636)
(3, 0.662264151)
(3.5, 0.704716981)
(4, 0.734433962)
(4.5, 0.758823529)
(5, 0.780882353)
(5.5, 0.794117647)
(6, 0.807352941)
(6.5, 0.820588235)
(7, 0.831122449)
(7.5, 0.840306122)
(8, 0.849489796)
(8.5, 0.858673469)
(9, 0.863265306)
(9.5, 0.867857143)
(10, 0.87244898)
(10.5, 0.880851064)
(11, 0.890425532)
(11.5, 0.895212766)
(12, 0.9)
(12.5, 0.904787234)
(13, 0.909574468)
(13.5, 0.914361702)
(14, 0.919148936)
(14.5, 0.919148936)
(15, 0.919148936)
(15.5, 0.92393617)
(16, 0.92393617)
(16.5, 0.93)
(17, 0.935)
(17.5, 0.935)
(18, 0.94)
(18.5, 0.945)
(19, 0.945)
(19.5, 0.95)
(20, 0.955)
(20.5, 0.955)
(21, 0.96)
(21.5, 0.96)
(22, 0.96)
(22.5, 0.96)
(23, 0.965)
(23.5, 0.965)
(24, 0.97)
(24.5, 0.97)
(25, 0.97)
};    

  \legend{}
  \addlegendentry{Inner MDS Code}
  \addlegendentry{Outer MDS Code}
  \addlegendentry{Inner+Outer MDS Code}
  \end{axis}
  \end{tikzpicture}
  \caption{$\epsilon=0.01$.}
  \label{fig4a}
  \end{subfigure} \hfill
\begin{subfigure}[b]{0.49\columnwidth}
\centering
\raisebox{0.2mm}{%
\begin{tikzpicture}[scale=0.54]
\begin{axis}[
    legend cell align={left},
    tick scale binop=\times,
    xlabel={Channel Error Rate $\epsilon$},
    ylabel={Maximum Code Rate $\rho^{\star}$},
    y label style={at={(axis description cs:0.07,.5)}, anchor=south},
    x label style={at={(axis description cs:0.5,0.022)}, anchor=north},
    grid=major,
    minor y tick num=1,
    grid=both,
    ytick = {0.70,0.75,0.80,0.85,0.90,0.95,1.0},
    xtick = {0.01,0.02,0.03,0.04,0.05,0.06,0.07,0.08,0.09,0.10},
    legend pos=south west,
    legend style={font=\small},
    xticklabel style={/pgf/number format/fixed},
    xmax=0.10,
    xmin=0.01,
    ymax=1.0,
    ymin=0.68,
    xmajorgrids=true,
    xminorgrids=true,
    ymajorgrids=true,
    yminorgrids=true,
    major grid style={black!30},
    minor grid style={dotted,black!50},
  ]
  \addplot+[color=blue, mark=triangle,mark size=1.5, mark options={solid, fill=none},line width=0.6pt] coordinates {
    (0.01, 0.957446808510638)
    (0.02, 0.957446808510638)
    (0.03, 0.957446808510638)
    (0.04, 0.918367346938776)
    (0.05, 0.918367346938776)
    (0.06, 0.918367346938776)
    (0.07, 0.918367346938776)
    (0.08, 0.882352941176471)
    (0.09, 0.882352941176471)
    (0.10, 0.849056603773585)
  };
  \addplot+[color=red, mark=o,mark size=1.5, mark options={solid, fill=none},line width=0.6pt] coordinates {
    (0.01, 0.918)
    (0.02, 0.872)
    (0.03, 0.852)
    (0.04, 0.877)
    (0.05, 0.868)
    (0.06, 0.853)
    (0.07, 0.832)
    (0.08, 0.840)
    (0.09, 0.817)
    (0.10, 0.819)
  };
  \addplot+[color=magenta, mark=x,mark size=1.5, mark options={solid, fill=none},line width=0.6pt] coordinates {
    (0.01, 0.878936170212766)
    (0.02, 0.834893617021277)
    (0.03, 0.815744680851064)
    (0.04, 0.805408163265306)
    (0.05, 0.797142857142857)
    (0.06, 0.783367346938776)
    (0.07, 0.764081632653061)
    (0.08, 0.741176470588235)
    (0.09, 0.720882352941176)
    (0.10, 0.695377358490566)
  };
  \legend{}
  \addlegendentry{Inner MDS Code}
  \addlegendentry{Outer MDS Code}
  \addlegendentry{Inner+Outer MDS Code}
\end{axis}
\end{tikzpicture}
}
  \caption{$R_{\text{all}}/K=10$.}
  \label{fig4b}
 \end{subfigure}
\caption{Maximum achievable code rate \( \rho^{\star} = \rho_{\text{in}}^{\star}\rho_{\text{out}}^{\star} \) (inner+outer) for reliable retrieval under the target threshold \( \delta_{\text{th}} = 10^{-6} \), together with the corresponding inner and outer code rates \( \rho_{\text{in}}^{\star} \) and \(\rho_{\text{out}}^{\star} \) resulting from the optimal redundancy allocation. The associated maximum information density in bits/NT is \mbox{$\Delta^{\star}=2\rho^{\star}$}. (a)~The information read depth \( R_{\text{all}} / K \) is varied for fixed channel error rate \( \epsilon = 0.01 \). (b)~The channel error rate \( \epsilon \) is varied for fixed information read depth \( R_{\text{all}} / K = 10 \). In both (a) and (b), the Dirichlet bias parameter is~\( \xi = 3 \).}
\label{fig4}
\end{figure}

Fig.~\ref{fig4b} illustrates the impact of the channel error rate $\epsilon$ on $\rho^{\star}$ for a fixed read depth $R_{\text{all}}/K=10$. As $\epsilon$ increases, the reliability constraint becomes more stringent and additional redundancy is required, leading to a decrease in the achievable code rate. The inner and outer rates shown in the figure reflect how the optimal redundancy allocation shifts as the channel becomes noisier, with a progressively larger portion of the total redundancy allocated to the inner code in the high-error regime.

Overall, these results highlight how the optimal redundancy allocation depends on the operating conditions. In low-noise or high-read regimes, the system can operate at high information densities by allocating limited redundancy through the outer code only. In contrast, in low-read or high-error regimes a larger portion of the total redundancy must be allocated to the inner code to maximize the achievable information density under reliability constraints.

The results in Fig.~\ref{fig3} and~\ref{fig4} highlight key trade-offs between design parameters that can be optimized to reduce sequencing or synthesis costs. To gain further insights into optimal DNA storage system design, it would be interesting to explore additional cost functions in future work, including ones that account for both synthesis and sequencing costs simultaneously.

\section{Beyond MDS Codes and QSC}
\label{sec5}

The theoretical framework we developed for evaluating the probability of successful data retrieval hinges on the interaction of four key components:  
(i) the post-consensus nucleotide and symbol error rates \( \epsilon_{(r)} \) and \( \epsilon'_{(r)} \) (Lemma~\ref{lemm1});  
(ii) the probabilities of success, miscorrection, and failure of the inner code, denoted \( \alpha_{(r)} \), \( \beta_{(r)} \), and \( \gamma_{(r)} \) (Lemma~\ref{lemma2});  
(iii) the conditional probability \(P_{\text{\normalfont succ} \mid  \boldsymbol{r}}\) of successful retrieval given a read profile \( \boldsymbol{r} \), which follows from the retrieval condition imposed by the outer code~(Theorem~\ref{thm1}); and  
(iv) a lower bound on the overall retrieval success probability \( P_{\text{\normalfont succ}} \) (Lemma~\ref{LB1} and Theorem~\ref{thm2}). 

Each of these components captures a different dimension of the system: the error model and consensus process (i), the structure and decoding behavior of the inner code (ii), the properties of the outer code (iii), and the read distribution under the channel model (iv). Owing to this modular structure, the framework remains applicable under different combinations of inner and outer codes as well as under alternative error models, provided each component is adapted accordingly.

\subsection{Asymmetric Substitutions}
A first and straightforward extension arises when the i.i.d.\ substitution assumption is maintained but the channel is asymmetric. In the baseline quaternary symmetric channel (QSC), each nucleotide is substituted by any of the three alternatives with equal probability \( \epsilon/3 \)~\eqref{QSC}. Under an asymmetric model, these transition probabilities differ across substitutions. In this case, Lemma~\ref{lemm1} remains valid with the same multinomial sum over \( (\kappa_1, \kappa_2, \kappa_3, \kappa_4) \in \mathcal{K}_{(r)} \), but the factors \( (1-\epsilon)^{\kappa_1} (\epsilon/3)^{r-\kappa_1} \) must be replaced by products of the appropriate per-base transition probabilities from the channel transition matrix, i.e., \( \prod_{i=1}^4 {\epsilon}_{b\to b_i}^{\kappa_i} \) conditioned on the true base~\(b\in \{\mathsf{A,C,G,T}\}\) (and averaged over \(b\) if the base prior is not uniform). Consequently,~\( \epsilon'_{(r)} \) follows by symbol grouping, and the remainder of the framework is unchanged.

\subsection{Inner Codes Beyond MDS}
Changing the inner code affects the second component of the framework (Lemma~\ref{lemma2}) because the probabilities of successful decoding, miscorrection, and failure depend on the chosen code and its decoding algorithm. When any $q$-ary block code of length \( n' \) and minimum Hamming distance \( d_{\min} \) is used as the inner code, the probability of successful decoding under bounded-distance decoding (BDD) retains the standard binomial-tail form
\[
\alpha_{(r)} = F(t; n', \epsilon'_{(r)}),
\]
where \( t = \lfloor (d_{\min} - 1)/2 \rfloor \) is the unique decoding radius and \( \epsilon'_{(r)} \) is the post-consensus {\em symbol} error rate. The latter follows from the post-consensus {\em nucleotide} error rate \( \epsilon_{(r)} \) (Lemma~\ref{lemm1}) and the number of bits per symbol \( m = \log_2 q \). Since each nucleotide encodes two bits, a symbol spans \( m/2 \) nucleotides, yielding $\epsilon'_{(r)} = 1 - \left( 1 - \epsilon_{(r)} \right)^{m/2}$, which reduces to \( \epsilon'_{(r)} = \epsilon_{(r)} \) in the case of a quaternary code (\( m = \log_2 4 = 2 \)). 

Quantifying the probabilities of miscorrection \( \beta_{(r)} \) and decoding failure \( \gamma_{(r)} \) is generally more challenging than determining $\alpha_{(r)}$, as they depend on the behavior of the code beyond its unique decoding radius and the specific code structure. The distinction between these two events is nevertheless essential since miscorrections propagate as substitutions to the outer decoder, whereas failures appear as erasures. For linear block codes decoded via BDD, the miscorrection and failure probabilities can be derived from the weight distribution of the code~\cite{mceliece2003decoder}. For linear MDS codes, these probabilities admit exact closed-form expressions (as mentioned in Lemma~\ref{lemma2}), and similar analyses extend to other algebraic codes such as BCH codes~\cite{miao2025errorratebinarybch}.

If the error correction capability of the inner code is not characterized by its minimum distance (e.g., convolutional codes), or if decoding relies on probabilistic algorithms such as belief propagation (e.g., LDPC codes) rather than BDD, exact closed-form expressions for \( \alpha_{(r)} \), \( \beta_{(r)} \), and \( \gamma_{(r)} \) are typically unavailable. In such cases, these quantities can be approximated or bounded analytically, or estimated numerically (e.g., via Monte Carlo simulation), and fed into the retrieval analysis without changing the framework. For conservative analytical bounds, one may also assume that all decoding outcomes beyond the unique decoding radius (when well-defined) result in miscorrections. While this likely leads to a loose lower bound on \( P_{\text{\normalfont succ} \mid \boldsymbol{r}} \), it preserves the tractability of the theoretical analysis.

\subsection{Outer Codes Beyond MDS}
The third component, \(P_{\text{\normalfont succ} \mid  \boldsymbol{r}}\), depends on the retrieval condition imposed by the outer code. As discussed in Section~\ref{sec3b}, a sufficient condition for successful retrieval for an \((N,K)\) outer MDS code is
$$
e^{\text{era}} + 2e^{\text{sub}} \leq N - K,
$$
where \( e^{\mathrm{era}} \) and \( e^{\mathrm{sub}} \) denote the numbers of sequence erasures and sequences containing substitutions, respectively. Here, erasures arise from decoding failures (with probability $\gamma$) and substitutions from miscorrections (with probability $\beta$) at the level of the inner code.

More generally, any outer block code with minimum Hamming distance $d_{\min}$ can correct all combinations of \( e^{\mathrm{era}} \) erasures and \( e^{\mathrm{sub}} \) substitutions provided that 
$$
e^{\text{era}} + 2e^{\text{sub}} \leq d_{\min} - 1.
$$
MDS codes are optimal in this regard since they achieve the Singleton bound with equality, $d_{\min}=N-K+1$, and therefore correct the maximum possible number of erasures and substitutions. This property makes them a natural choice for the outer code in DNA storage. However, Reed-Solomon codes and other MDS codes can become computationally demanding to decode as the blocklength grows. In practice, outer codes with lower decoding complexity but slightly reduced error correction capability, such as LDPC codes, may be adopted when decoding efficiency is a priority. In such cases, the retrieval condition must be reformulated.

In the case of LDPC codes, modified belief-propagation decoders can jointly handle combinations of erasures and substitutions. However, their performance is not captured by simple distance-based conditions but rather by iterative decoding thresholds that characterize the point below which decoding converges with high probability. These thresholds depend on the specific code ensemble, graph structure, and channel parameters, and they can be evaluated analytically (e.g., via density evolution) or estimated empirically via numerical simulations.

Fountain codes, such as Raptor codes, are also an attractive choice for outer codes due to their {\em rateless} nature: they can generate a virtually unlimited number of encoded sequences, enabling the selection of only those that satisfy specific biochemical constraints for synthesis while discarding the rest, a feature that has been exploited in wet-lab experiments~\cite{erlich2017dna}. They further offer linear-time encoding and decoding, making them highly scalable for large-scale storage. However, their main limitation is that they are designed to correct only erasures and require a decoding overhead. Namely, successful retrieval typically requires that \( K(1+\nu) \) encoded sequences are correctly received, where $\nu>0$ denotes the overhead. More critically, fountain codes can fail catastrophically if erroneous sequences are included in the decoding set, since even a few incorrect sequences can render the underlying linear system unsolvable. As a result, the condition for successful retrieval takes the approximate form
\begin{equation*}
e^{\mathrm{sub}} \approx 0 \quad \text{and} \quad e^{\mathrm{era}} \leq N - K(1+\nu),
\end{equation*}
which is significantly more restrictive than that of MDS codes. This highlights the importance of pairing fountain codes with additional inner-code detection mechanisms such as cyclic redundancy checks (CRCs), which can convert potential miscorrections into decoding failures.

\subsection{Deletions and Insertions}
When the error model includes general edits, i.e., deletions and insertions in addition to substitutions, the principal change occurs in the post-consensus error rate $\epsilon_{(r)}$ (first component), which must now account for an alignment step preceding consensus, since reads are no longer naturally aligned. This presents a major challenge because, beyond determining the error rate itself, it is no longer clear that the statistics of the residual errors in the consensus sequence will remain i.i.d.\ as in the substitution-only case, even if all edits are i.i.d. at the base level. 

Furthermore, all these considerations depend strongly on the multiple sequence alignment (MSA) algorithm used, and such algorithms are often difficult to analyze. As a result, a theoretical characterization of the post-consensus error rate and its statistics may be intractable. A more practical approach is to estimate these quantities empirically for a fixed MSA+consensus algorithm as a function of the number of reads, and then use the results in other components of the~framework.

As for the inner code (second component), if a quaternary block code with minimum Levenshtein distance \( d_L \) is used, the probability of successful decoding \( \alpha_{(r)} \) under an edit channel is determined by the probability that the number of residual edits does not exceed \( t_L = \lfloor (d_L - 1)/2 \rfloor \). This probability depends on the statistics of the residual errors. If these errors can be approximated as i.i.d., then \( \alpha_{(r)} \) retains the standard binomial-tail form, analogous to the substitution-only case:
\[
\alpha_{(r)} = F(t_L; n', \epsilon_{(r)}),
\]
where \( \epsilon_{(r)} \) now accounts for all three types of edits. If the residual errors follow different statistics, the functional form of \( \alpha_{(r)} \) will change accordingly, but it remains computable once the distribution of the total number of errors is known. As an example, for the quaternary single-edit-correcting code introduced in~\cite{cai2021correcting}, \( \alpha_{(r)} \) can be evaluated directly for \( t_L = 1 \). For other edit-correcting codes with probabilistic decoding, analytical quantifications or closed-form expressions for \( \alpha_{(r)} \) are generally difficult to obtain. An exception is the GC+ code~\cite{hanna2025gccodesystematicshort}, for which analytical bounds exist and can be incorporated into the analysis. As in the substitution case, the distinction between \( \beta_{(r)} \) and \( \gamma_{(r)} \) depends more strongly on the structure of the code. Importantly, the third and fourth components of the framework remain unaffected by the presence of edit errors, as they depend only on the numbers of correct, erroneous, and missing sequences rather than on the nature of the errors.


\section{Conclusion}
This paper presented a theoretical framework for analyzing the reliability of information retrieval from MDS coded data in DNA storage. The framework links four components: the post-consensus error rates, the inner-code outcome probabilities, an outer-code retrieval condition, and a lower bound on the overall success probability. The bound leverages summary statistics of the read profile and admits efficient evaluation, which we used to study design trade-offs and to solve representative optimization problems on minimizing sequencing and synthesis costs in DNA data storage systems. While our analysis focused on MDS inner and outer codes and i.i.d. substitution errors, we outlined how the framework can be potentially applied to broader scenarios, including alternative inner and outer codes, and channels involving asymmetric substitutions or edit errors. 

Throughout this work, consensus formation, inner decoding, and outer decoding were treated as distinct components, each analyzed independently and then combined into an end-to-end reliability expression, which is the most common practice in the literature. However, this separation is not the only possible design choice. For example, consensus formation and inner decoding could be combined into a single reconstruction step, or inner and outer codes could be co-designed as a single two-dimensional code.  A natural future direction is to analyze jointly designed approaches, in which these components are optimized together rather than separately, and to theoretically quantify the achievable gains in reliability and cost-efficiency while keeping the overall system practical and scalable.

\section*{Acknowledgment} 
This work was supported by the French government through the France 2030 investment plan managed by the National Research Agency (ANR), as part of the Initiative of Excellence Université Côte d’Azur under reference number ANR-15-IDEX-01, and by ANR under reference number ANR-22-CPJ2-0054-01.

\appendices
\section{Proof of Lemma~\ref{lemm1}}
Let $\tilde{\Rv{Y}}^{1}, \tilde{\Rv{Y}}^{2}, \ldots, \tilde{\Rv{Y}}^{r}\in \Sigma^{n/2}$ be the $r\in \mathbb{N}$ noisy copies generated by the channel corresponding to a given DNA sequence $\tilde{\boldsymbol{x}} \in \Sigma^{n/2}$, where $\Sigma = \{\mathsf{A,C,G,T}\}$. For each position $i \in [n/2]$, the consensus nucleotide $\tilde{\rv{C}}_i\in \Sigma$ is chosen as the most frequent one among $\tilde{\rv{Y}}_i^{1},\tilde{\rv{Y}}_i^{2}, \ldots, \tilde{\rv{Y}}_i^{r}$, with ties broken uniformly at random. Since the QSC introduces i.i.d. errors and the consensus mechanism processes each nucleotide position independently, the errors in the consensus sequence also remain i.i.d., with error probability 
$$\epsilon_{(r)}=\Pr(\tilde{\rv{C}}_i \neq \tilde{x}_i), \quad \text{for any } i \in [n/2].$$
Due to the symmetry across positions, we drop the subscript $i$ in the rest of the proof. Let $(\rv{K}_1,\rv{K}_2,\rv{K}_3,\rv{K}_4)$ be the random vector defined as $$(\rv{K}_1,\rv{K}_2,\rv{K}_3,\rv{K}_4)\in \left\{ \boldsymbol{\kappa} \in \mathbb{N}^4 \ : \ \kappa_1 + \kappa_2 + \kappa_3 + \kappa_4 = r \right\},$$  where $$\rv{K}_1=\sum_{\ell=1}^r \mathds{1}_{\{\tilde{\rv{Y}}^{\ell} = \tilde{x}\}}$$ 
counts the number of copies in which the nucleotide $\tilde{x}$ is retained with no error, and $\rv{K}_2$,$\rv{K}_3$, and $\rv{K}_4$ count the number of copies in which $\tilde{x}$ is substituted to each of the other three nucleotides in $\Sigma \setminus \tilde{x}$. The vector $(\rv{K}_1,\rv{K}_2,\rv{K}_3,\rv{K}_4)$ follows a multinomial distribution with parameter $r$ (number of trials) and probability vector $(1-\epsilon,\frac{\epsilon}{3},\frac{\epsilon}{3},\frac{\epsilon}{3})$, where $\epsilon$ is the QSC error rate. For a realization $\boldsymbol{\kappa} = (\kappa_1, \kappa_2, \kappa_3, \kappa_4)$, the probability of correctly recovering $\tilde{x}$ via majority voting (with ties broken uniformly) is
$$\Pr(\tilde{\rv{C}} = \tilde{x} \mid \kappa_1,\kappa_2,\kappa_3,\kappa_4) = \frac{\mathds{1}_{\{\kappa_1 \geq \kappa_2,\kappa_3,\kappa_4\}}}{\sum_{i=1}^4 \mathds{1}_{\{\kappa_i = \kappa_1\}}}  ,
$$
where the denominator \mbox{$\omega_{(\boldsymbol{\kappa})} = \sum_{i=1}^4 \mathds{1}_{\{\kappa_i = \kappa_1\}}$} counts the number of ties. Applying the law of total probability,
\begin{align*}
  \Pr(\tilde{\rv{C}} = \tilde{x}) &=  \hspace{-0.75cm} \sum_{\substack{\boldsymbol{\kappa}\in \mathbb{N}^4 \\ \kappa_1+\kappa_2+\kappa_3+\kappa_4=r}} \hspace{-0.75cm}  \Pr(\tilde{\rv{C}} = \tilde{x} \mid \kappa_1,\kappa_2,\kappa_3,\kappa_4) \Pr(\kappa_1,\kappa_2,\kappa_3,\kappa_4),\\
  &= \sum_{\boldsymbol{\kappa}\in \mathcal{K}_{(r)}} \frac{1}{\omega_{(\boldsymbol{\kappa})}} \binom{r}{\kappa_1,\kappa_2,\kappa_3,\kappa_4}(1-\epsilon)^{\kappa_1} \left(\frac{\epsilon}{3} \right)^{r-\kappa_1},
\end{align*}
where
$$\mathcal{K}_{(r)} = \bigg\{ \boldsymbol{\kappa} \in \mathbb{N}^4 \ : \ \sum_{i=1}^{4} \kappa_i = r, \kappa_1 \geq  \kappa_2,  \kappa_3,  \kappa_4 \bigg\}.$$
Thus, the post-consensus error rate is $$\epsilon_{(r)} = \Pr(\tilde{\rv{C}} \neq \tilde{x}) = 1 -   \Pr(\tilde{\rv{C}} = \tilde{x}).$$ Furthermore, since the errors in the consensus sequence are i.i.d. and each symbol in the inner code corresponds to $m/2$ nucleotides, the inner code symbol error rate is $$\epsilon'_{(r)} = 1- (\Pr(\tilde{\rv{C}} = \tilde{x}))^{\frac{m}{2}} = 1 - \left(1- \epsilon_{(r)}\right)^{\frac{m}{2}},$$ which concludes the proof.

\section{Proof of Lemma~\ref{lemma2}}
The inner \mbox{$(n' = \frac{n}{m}, k' = \frac{k}{m})$} MDS code, where $n' - k' = 2t$, can correct up to $t$ symbol substitution errors under bounded-distance decoding (BDD). The number of symbol errors in a codeword follows a binomial distribution with parameters $(n', \epsilon'_{(r)})$. Hence, the probability of successful decoding is
\[
\alpha_{(r)} = \sum_{i=0}^t \binom{n'}{i} (\epsilon'_{(r)})^{i} (1-\epsilon'_{(r)})^{n'-i} = F(t; n', \epsilon'_{(r)}).
\]

When the number of errors exceeds $t$, decoding may result in either a miscorrection or a decoding failure. Specifically, for the cases with \mbox{$i \in [t+1, n']$} symbol errors, a fraction $\eta_i \in (0,1)$ of these error patterns result in a miscorrection, while the remaining fraction $1 - \eta_i$ lead to a decoding failure. Exact closed-form expressions for $\eta_i$ applicable to all linear MDS codes (e.g., Reed-Solomon codes) are provided in~\cite{cheung1989more}. Consequently, the probabilities of miscorrection and decoding failure are given by
\[
\beta_{(r)} = \sum_{i=t+1}^{n'} \eta_i f(i; n', \epsilon'_{(r)}),
\]
\[
\gamma_{(r)} = \sum_{i=t+1}^{n'} (1 - \eta_i) f(i; n', \epsilon'_{(r)}),
\]
where $f$ denotes the PMF of a binomial distribution as defined in Section~\ref{sec2}.
Since these three events are mutually exclusive and collectively exhaustive, their probabilities sum to one.

\section{Proof of Theorem~\ref{thm1}}
Theorem~\ref{thm1} follows from Lemmas~\ref{lemm1} and~\ref{lemma2} combined with a standard property of MDS code, namely, the outer $(N, K)$ MDS code can correct any combination of $e^{\text{era}}$ erasures and $e^{\text{sub}}$ substitutions if \mbox{$e^{\text{era}} + 2e^{\text{sub}} \leq N - K$}. In our setting, $e^{\text{era}}$ and $e^{\text{sub}}$ correspond to the total number of decoding failures and miscorrections, respectively, arising from the decoding of the $N$ individual sequences using the inner MDS code. Equivalently, this condition can be written as $$(N-e_{\text{era}}-e_{\text{sub}})-e_{\text{sub}} \geq K,$$ where $N-e_{\text{era}}-e_{\text{sub}}$ counts the number of correctly decoded sequences (successful decoding), while $e_{\text{sub}}$ corresponds to the number of incorrectly decoded sequences (miscorrection). By definition, we have
$$\rv{S}_N{(\boldsymbol{r})} = \sum_{j=1}^N \rv{Z}_j{(r_j)} = (N-e_{\text{era}}-e_{\text{sub}})-e_{\text{sub}}.$$ Thus, $\rv{S}_N{(\boldsymbol{r})}\geq K$ is a sufficient condition for recovering all information symbols correctly. Therefore, the probability of successful retrieval satisfies $$P_{\text{\normalfont succ}\mid\boldsymbol{r}}\geq \Pr\left(\rv{S}_N{(\boldsymbol{r})}\geq K \right),$$ which holds with equality if $M=k$ as noted in Remark~\ref{rem:score_condition}.

Let $\boldsymbol{z}=(z_1,\ldots,z_N)\in \{-1,0,1\}^N$ be a realization of $(\rv{Z}_1{(r_1)}, \ldots, \rv{Z}_N{(r_N)})$. Since $\rv{Z}_1{(r_1)}, \ldots, \rv{Z}_N{(r_N)}$ are independent for a given read profile $\boldsymbol{r}$, we obtain
$$\Pr(\rv{S}_N{(\boldsymbol{r})}\geq K) = \sum_{\substack{\boldsymbol{z} \in \{-1,0,1\}^N \\ K\leq z_1+\ldots+z_N \leq N }} \hspace{-0.1cm} \prod_{j=1}^N \Pr\left(\rv{Z}_j{(r_j)} = z_j \right),$$
where
\begin{align*}
    \Pr(\rv{Z}_j{(r_j)} = 1) &= \alpha_{(r_j)}, \\
    \Pr(\rv{Z}_j{(r_j)} = -1) &= \beta_{(r_j)}, \\
    \Pr(\rv{Z}_j{(r_j)} = 0) &= \gamma_{(r_j)}.
\end{align*}
Alternatively, as shown in~\eqref{eqP0}, an equivalent expression can be derived by summing over subsets $\mathcal{A},\mathcal{B} \subseteq [N]$, where $\mathcal{A}$ indexes the correctly decoded sequences and $\mathcal{B}$ indexes the incorrectly decoded ones, with $\mathcal{A} \cap \mathcal{B} = \varnothing$ and  $|\mathcal{A}| - |\mathcal{B}|\geq K$.

\section{Proof of Corollary~\ref{corr4}}
From Theorem~\ref{thm1} we have
$$P_{\text{\normalfont succ}\mid r}\geq \Pr\left(\rv{S}_N{(r)}\geq K \right)=\sum_{s=K}^N \Pr\left(\rv{S}_N{(r)}= s \right).$$
Define the following random variables
$$N^+ \triangleq \sum_{j=1}^N \mathds{1}_{\{\rv{Z}_j(r) = +1\}}, \quad N^- \triangleq\sum_{j=1}^N \mathds{1}_{\{\rv{Z}_j(r) = -1\}},$$
and $N^0 \triangleq N - N^+ - N^-$. The vector $(N^+,N^-,N^0)$ follows a multinomial distribution with parameters $N$ and \mbox{$(\alpha_{(r)},\beta_{(r)},\gamma_{(r)})$}. The PMF of \mbox{$\rv{S}_N{(r)}\in [-N, N]$} is given by 
\begin{align*}
 \Pr\left(\rv{S}= s \right) &= \Pr(N^{+}-N^{-}=s), \\
 &= \sum_{i} \Pr(N^+=i+s, N^- = i, N^0 = N-2i-s), \\
 &= \sum_{i}  \binom{N}{i+s,i,N-2i-s} \alpha_{(r)}^{i+s} \beta_{(r)}^{i} \gamma_{(r)}^{N-2i-s}.
\end{align*}
The limits of the summation above are determined by the following conditions: 
$$N^+ + N^- \leq N \Longrightarrow 2i+s\leq N \Longrightarrow i\leq \left\lfloor \frac{N-s}{2} \right\rfloor,$$
$$N^+ \geq 0 \Longrightarrow i\geq -s, \quad N^-\geq 0 \Longrightarrow i\geq 0.$$
Therefore, $i\geq \max \{-s,0\}$ and $i\leq \lfloor (N-s)/2 \rfloor$, which concludes the proof.

\section{Proof of Corollary~\ref{corr5}}
\noindent Recall that each $\rv{Z}_j(r_j)\in\{-1,0,1\}$ has probabilities 
$\alpha_{(r_j)},\;\beta_{(r_j)},\;\gamma_{(r_j)}$ 
of taking values $\{+1,\,-1,\,0\}$, respectively. Hence, for $j\in [N]$, 
\begin{align*}
    \mu_{(r_j)} &\triangleq \mathbb{E}[\rv{Z}_j(r_j)] =  \alpha_{(r_j)}-\beta_{(r_j)}, \\
    \sigma^2_{(r_j)} &\triangleq \mathbb{V}[\rv{Z}_j(r_j)] = \alpha_{(r_j)}+\beta_{(r_j)}-\left(\alpha_{(r_j)}-\beta_{(r_j)}\right)^2.
\end{align*}
Thus, by linearity of expectation,
\begin{align*}
      \mu_{(\boldsymbol{r})} &= \mathbb{E}[\rv{S}_N(\boldsymbol{r})] = \sum_{j=1}^N \mu_{(r_j)},
\end{align*}
and since $\rv{Z}_{1}{(r_1)},\ldots,\rv{Z}_{N}{(r_N)}$ are independent,
$$\sigma_{(\boldsymbol{r})}^2 = \mathbb{V}[\rv{S}_N(\boldsymbol{r})] = \sum_{j=1}^N \sigma^2_{(r_j)}.$$
To show that $\rv{S}_N(\boldsymbol{r})$ is approximately Gaussian for large $N$, we use Lyapunov’s CLT, which is a generalization of the classical CLT for the sum of independent, but not necessarily identically distributed, random variables. Lyapunov's condition says that if there is some $\delta>0$ such that
\[
\lim_{N \to \infty} \frac{1}{\sigma_{(\boldsymbol{r})}^{2+\delta}} \sum_{j=1}^{N} \mathbb{E}\left[ \big|\rv{Z}_j(r_j) - \mu{(r_j)}\big|^{2+\delta} \right] = 0,
\]
then $(\rv{S}_N(\boldsymbol{r}) - \mu_{(\boldsymbol{r})})/\sigma_{(\boldsymbol{r})}$ converges in distribution to a standard Gaussian random variable. Since $\rv{Z}_{j}{(r_j)}\in\{-1,0,1\}$ is bounded between $-1$ and $1$ for all $j$, then it holds that
$$\big|\rv{Z}_j(r_j) - \mu{(r_j)}\big| \leq 2~~\Longrightarrow~~\mathbb{E}\left[ \big|\rv{Z}_j(r_j) - \mu{(r_j)}\big|^{2+\delta}\right] \leq 2^{2+\delta},$$
and hence 
$$\sum_{j=1}^{N} \mathbb{E}\left[ \big|\rv{Z}_j(r_j) - \mu{(r_j)}\big|^{2+\delta} \right] \leq N 2^{2+\delta}.$$
Thus, Lyapunov's condition reduces to 
\[
\lim_{N \to \infty} \frac{2^{2+\delta}N}{\sigma_{(\boldsymbol{r})}^{2+\delta}} = \lim_{N \to \infty} \frac{N}{\sigma_{(\boldsymbol{r})}^{2+\delta}} = 0,
\]
which requires 
$\sigma_{(\boldsymbol{r})}^{2+\delta} = \left[\textstyle \sum_{j=1}^N \sigma^2_{(r_j)}\right]^{1+\frac{\delta}{2}}$
to be superlinear in $N$ for some $\delta>0$, i.e., $\sigma_{(\boldsymbol{r})}^{2+\delta}=\omega(N)$. It follows from Lemma~\ref{lemma2} that $\sigma^2_{(r_j)}=0$ if $r_j=0$, and $\sigma^2_{(r_j)}>0$ if $r_j\geq 1$. Consequently, there must be a sufficient number of non-zeros in \mbox{$\boldsymbol{r}=(r_1,\ldots,r_N)$} so that the condition is satisfied. If $\|\boldsymbol{r} \|_0\geq N^{\theta}$ for some constant $\theta\in (0,1]$, then there exists a constant $c>0$ such that
$$\sigma_{(\boldsymbol{r})}^{2} = \sum_{j=1}^N \sigma^2_{(r_j)}\geq  c\| \boldsymbol{r}\|_0 \geq cN^{\theta}.$$
Therefore,
$$ \lim_{N \to \infty} \frac{N}{\sigma_{(\boldsymbol{r})}^{2+\delta}} \leq \lim_{N \to \infty} \frac{N}{\left[ cN^{\theta} \right]^{1+\frac{\delta}{2}}} = \lim_{N \to \infty} N^{1-\theta(1+\frac{\delta}{2})}.$$
This limit is zero when 
$$1-\theta\left(1+\frac{\delta}{2}\right)<0~~\Longleftrightarrow~~\delta > 2\left(\frac{1}{\theta} - 1 \right).$$
Since $\frac{1}{\theta}\geq 1$ for $\theta\in (0,1]$, we can always choose a positive $\delta$ large enough to satisfy $\delta>2\left(\frac{1}{\theta} - 1 \right)$. Lyapunov’s condition is therefore satisfied, implying that for large $N$ we have
\begin{align*}
    \Pr(\rv{S}_N(\boldsymbol{r})\geq K) &= \Pr\left(\frac{\rv{S}_N(\boldsymbol{r}) - \mu_{(\boldsymbol{r})}}{\sigma_{(\boldsymbol{r})}}\geq \frac{K-\mu_{(\boldsymbol{r})}}{\sigma_{(\boldsymbol{r})}}\right), \\
    &\approx 1 - \Phi\left(\frac{K-0.5-\mu_{(\boldsymbol{r})}}{\sigma_{(\boldsymbol{r})}}\right), \\
    &= \Phi\left(\frac{\mu_{(\boldsymbol{r})}-K+0.5}{\sigma_{(\boldsymbol{r})}}\right),
\end{align*}
where $\Phi$ is the CDF of a standard Gaussian distribution and the shift by 0.5 is for continuity correction. The following bound on the approximation error can be thus obtained from the Berry-Esseen inequality applied to Lyapunov's CLT~\cite{shevtsova2010improvement}:
$$\left \lvert  \Pr(\rv{S}_N(\boldsymbol{r})\geq K) -  \Phi\left(\frac{\mu_{(\boldsymbol{r})}-K+0.5}{\sigma_{(\boldsymbol{r})}}\right) \right \rvert \leq  \frac{C \zeta_{(\boldsymbol{r})}}{ \sqrt{N} \sigma_{(\boldsymbol{r})}^{3/2}} ,$$
where $C\leq 0.5606$ is a constant, and $\zeta_{\boldsymbol{r}}$ denotes the sum of the third absolute central moments of $\rv{Z}_{1}{(r_1)},\ldots,\rv{Z}_{N}{(r_N)}$, i.e.,
\begin{align*}
    \zeta_{(\boldsymbol{r})} &= \sum_{j=1}^N \mathbb{E} \left[\big|\rv{Z}_j(r_j) - \mu_{(r_j)} \big|^3 \right] \\
    &= \sum_{j=1}^N \alpha_{(r_j)} \big|1 - \mu_{(r_j)} \big|^3 + \beta_{(r_j)} \big|1 + \mu_{(r_j)} \big|^3 + \gamma_{(r_j)} \big|\mu_{(r_j)} \big|^3.
\end{align*}

\section{Proofs of Lemma~\ref{LB1} and Theorem~\ref{thm2}}
\noindent Recall that
\begin{equation*}
P_{\text{succ}\mid \boldsymbol{r}}\geq \Pr\left(\rv{S}_N{(\boldsymbol{r})}\geq K \right),
\end{equation*}
where $\rv{S}_N{(\boldsymbol{r})} =\rv{Z}_1{(r_1)}+\rv{Z}_2{(r_2)}+\ldots+\rv{Z}_N{(r_N)}$. For a given read profile $\boldsymbol{r}=(r_1,\ldots,r_N)$, let $\Tilde{\boldsymbol{r}}=(\Tilde{r}_1,\Tilde{r}_2,\ldots,\Tilde{r}_N)$ denote the elements of $\boldsymbol{r}$ sorted in increasing order. By symmetry across sequences, we have
\begin{equation*}
   \rv{S}_N{(\boldsymbol{r})} = \rv{S}_N{(\Tilde{\boldsymbol{r}})} =\rv{Z}_1{(\Tilde{r}_1)}+\rv{Z}_2{(\Tilde{r}_2)}+\ldots+\rv{Z}_N{(\Tilde{r}_N)}.
\end{equation*}
For any $r' \in [R_{\text{all}}]$, $\tilde{h}_{r'}=h_0+h_1+\ldots+h_{r'}$ represents the number of sequences that are read at most $r'$ times. Given $r'$, $\tilde{h}_{r'}$, $h_0$, and $\Tilde{\boldsymbol{r}}$, consider the random variables
\begin{equation*}
    \begin{aligned}
        \rv{S}_{\text{low}} &= \sum_{j=1}^{\tilde{h}_{r'}} \rv{Z}_j(\tilde{r}_j), \\
        \rv{S}_{\text{high}} &= \sum_{j=\tilde{h}_{r'}+1}^{N} \rv{Z}_j(\tilde{r}_j),
    \end{aligned}
    \qquad
    \begin{aligned}
        \rv{S}'_{\text{low}} &= \sum_{j=h_0+1}^{\tilde{h}_{r'}} \rv{Z}_j(1), \\
        \rv{S}'_{\text{high}} &= \sum_{j=\tilde{h}_{r'}+1}^{N} \rv{Z}_j(r'+1),
    \end{aligned}
\end{equation*}
where $\rv{S}_{\text{low}}+\rv{S}_{\text{high}}={S}_N{(\boldsymbol{r})}$, and $${S}'_{\text{low}}\subseteq [-(\tilde{h}_{r'}-h_0),\tilde{h}_{r'}-h_0],~~~ {S}'_{\text{high}}\subseteq [-(N-\tilde{h}_{r'}),N-\tilde{h}_{r'}].$$ It follows from Lemmas~\ref{lemm1} and \ref{lemma2} that $\alpha_{(r)}=\Pr(\rv{Z}(r)=+1)$ is an increasing function of $r$. Thus, one can easily show that
$$\Pr\left(\rv{S}_{\text{low}}+\rv{S}_{\text{high}}\geq K \right) \geq \Pr \left(\rv{S}'_{\text{low}} + \rv{S}'_{\text{high}} \geq K \right).$$ 

Furthermore, by definition, $\rv{S}'_{\text{low}}$ and $\rv{S}'_{\text{high}}$ are sums of inner-decoding indicator variables corresponding to subsets of sequences with uniform read counts (one read per sequence for $\rv{S}'_{\text{low}}$ and $r'+1$ reads for  $\rv{S}'_{\text{high}}$). Hence, the summands in each case are independent and identically distributed, and Corollary~\ref{corr4} applies. The PMFs of $\rv{S}'_{\text{low}}$ and $\rv{S}'_{\text{high}}$ can thus be obtained from Corollary~\ref{corr4} by substituting $(N,r)$ with $(\tilde{h}_{r'}-h_0,1)$ and $(N-\tilde{h}_{r'},r'+1)$, respectively. Therefore,
\begin{align*}
   P_{\text{succ} \mid  \boldsymbol{r}} &\geq  \Pr\left(\rv{S}_N{(\boldsymbol{r})}\geq K \right), \\
   &= \Pr\left(\rv{S}_{\text{low}}+\rv{S}_{\text{high}}\geq K \right), \\
   &\geq \Pr \left(\rv{S}'_{\text{low}} + \rv{S}'_{\text{high}} \geq K \right), \\
   &\geq \sum_{s\in \mathcal{S}}  \Pr (\rv{S}'_{\text{high}} \geq K-s \mid \rv{S}'_{\text{low}}=s ) \Pr(\rv{S}'_{\text{low}} = s),~~~~~~~~~~~~
\end{align*}
where the last inequality follows from applying the law of total probability by marginalizing over $\rv{S}'_{\text{low}}$ for any subset \mbox{$\mathcal{S} \subseteq [-(\tilde{h}_{r'}-h_0),\tilde{h}_{r'}-h_0]$}. This bound is of the form \mbox{$P_{\text{succ} \mid  \boldsymbol{r}}\geq P_{\text{succ} \mid  h_0,\tilde{h}_{r'}}$}, which depends on $\boldsymbol{r}$ only through~$h_0$ (the number of sequences with zero reads) and $\tilde{h}_{r'}$ (the number of sequences with at most $r'$ reads). Consequently, the proof of Theorem~\ref{thm2} follows by applying the law of total probability over the pair $(\rv{H}_0, \tilde{\rv{H}}_{r'})$, using their joint PMF derived from $\eqref{eqRecc}$ (for $j=N$), along with the lower bound given in Lemma~\ref{LB1}. %

\bibliographystyle{IEEEtran}
\bibliography{Refs}

\begin{thebibliography}{10}
\providecommand{\url}[1]{#1}
\csname url@samestyle\endcsname
\providecommand{\newblock}{\relax}
\providecommand{\bibinfo}[2]{#2}
\providecommand{\BIBentrySTDinterwordspacing}{\spaceskip=0pt\relax}
\providecommand{\BIBentryALTinterwordstretchfactor}{4}
\providecommand{\BIBentryALTinterwordspacing}{\spaceskip=\fontdimen2\font plus
\BIBentryALTinterwordstretchfactor\fontdimen3\font minus
  \fontdimen4\font\relax}
\providecommand{\BIBforeignlanguage}[2]{{%
\expandafter\ifx\csname l@#1\endcsname\relax
\typeout{** WARNING: IEEEtran.bst: No hyphenation pattern has been}%
\typeout{** loaded for the language `#1'. Using the pattern for}%
\typeout{** the default language instead.}%
\else
\language=\csname l@#1\endcsname
\fi
#2}}
\providecommand{\BIBdecl}{\relax}
\BIBdecl

\bibitem{11195362}
S.~Kas~Hanna, ``On the reliability of information retrieval from {MDS} coded
  data in {DNA} storage,'' in \emph{2025 IEEE International Symposium on
  Information Theory (ISIT)}, 2025, pp. 1--6.

\bibitem{rydning2022worldwide}
J.~Rydning, ``Worldwide idc global datasphere forecast, 2022--2026: enterprise
  organizations driving most of the data growth,'' \emph{International Data
  Corporation (IDC)}, 2022.

\bibitem{church2012next}
G.~M. Church, Y.~Gao, and S.~Kosuri, ``Next-generation digital information
  storage in {DNA},'' \emph{Science}, vol. 337, no. 6102, pp. 1628--1628, 2012.

\bibitem{grass2015robust}
R.~N. Grass, R.~Heckel, M.~Puddu, D.~Paunescu, and W.~J. Stark, ``Robust
  chemical preservation of digital information on {DNA} in silica with
  error-correcting codes,'' \emph{Angewandte Chemie International Edition},
  vol.~54, no.~8, pp. 2552--2555, 2015.

\bibitem{ceze2019molecular}
L.~Ceze, J.~Nivala, and K.~Strauss, ``Molecular digital data storage using
  {DNA},'' \emph{Nature Reviews Genetics}, vol.~20, no.~8, pp. 456--466, 2019.

\bibitem{yazdi2015dna}
S.~H.~T. Yazdi, H.~M. Kiah, E.~Garcia-Ruiz, J.~Ma, H.~Zhao, and O.~Milenkovic,
  ``{DNA}-based storage: Trends and methods,'' \emph{IEEE Transactions on
  Molecular, Biological and Multi-Scale Communications}, vol.~1, no.~3, pp.
  230--248, 2015.

\bibitem{shomorony2022information}
I.~Shomorony, R.~Heckel \emph{et~al.}, ``Information-theoretic foundations of
  {DNA} data storage,'' \emph{Foundations and Trends{\textregistered} in
  Communications and Information Theory}, vol.~19, no.~1, pp. 1--106, 2022.

\bibitem{heckel2019characterization}
R.~Heckel, G.~Mikutis, and R.~N. Grass, ``A characterization of the {DNA} data
  storage channel,'' \emph{Scientific reports}, vol.~9, no.~1, p. 9663, 2019.

\bibitem{meiser2020reading}
L.~C. Meiser, P.~L. Antkowiak, J.~Koch, W.~D. Chen, A.~X. Kohll, W.~J. Stark,
  R.~Heckel, and R.~N. Grass, ``Reading and writing digital data in {DNA},''
  \emph{Nature protocols}, vol.~15, no.~1, pp. 86--101, 2020.

\bibitem{gimpel2023digital}
A.~L. Gimpel, W.~J. Stark, R.~Heckel, and R.~N. Grass, ``A digital twin for
  {DNA} data storage based on comprehensive quantification of errors and
  biases,'' \emph{Nature Communications}, vol.~14, no.~1, p. 6026, 2023.

\bibitem{blawat2016forward}
M.~Blawat, K.~Gaedke, I.~Huetter, X.-M. Chen, B.~Turczyk, S.~Inverso, B.~W.
  Pruitt, and G.~M. Church, ``Forward error correction for {DNA} data
  storage,'' \emph{Procedia Computer Science}, vol.~80, pp. 1011--1022, 2016.

\bibitem{erlich2017dna}
Y.~Erlich and D.~Zielinski, ``{DNA} fountain enables a robust and efficient
  storage architecture,'' \emph{science}, vol. 355, no. 6328, pp. 950--954,
  2017.

\bibitem{yazdi2017portable}
S.~H.~T. Yazdi, R.~Gabrys, and O.~Milenkovic, ``Portable and error-free
  {DNA}-based data storage,'' \emph{Scientific reports}, vol.~7, no.~1, p.
  5011, 2017.

\bibitem{organick2018random}
L.~Organick, S.~D. Ang, Y.-J. Chen, R.~Lopez, S.~Yekhanin, K.~Makarychev, M.~Z.
  Racz, G.~Kamath, P.~Gopalan, B.~Nguyen \emph{et~al.}, ``Random access in
  large-scale {DNA} data storage,'' \emph{Nature biotechnology}, vol.~36,
  no.~3, pp. 242--248, 2018.

\bibitem{antkowiak2020low}
P.~L. Antkowiak, J.~Lietard, M.~Z. Darestani, M.~M. Somoza, W.~J. Stark,
  R.~Heckel, and R.~N. Grass, ``Low cost {DNA} data storage using
  photolithographic synthesis and advanced information reconstruction and error
  correction,'' \emph{Nature communications}, vol.~11, no.~1, p. 5345, 2020.

\bibitem{chandak2020overcoming}
S.~Chandak, J.~Neu, K.~Tatwawadi, J.~Mardia, B.~Lau, M.~Kubit, R.~Hulett,
  P.~Griffin, M.~Wootters, T.~Weissman \emph{et~al.}, ``Overcoming high
  nanopore basecaller error rates for {DNA} storage via basecaller-decoder
  integration and convolutional codes,'' in \emph{ICASSP 2020-2020 IEEE
  International Conference on Acoustics, Speech and Signal Processing
  (ICASSP)}.\hskip 1em plus 0.5em minus 0.4em\relax IEEE, 2020, pp. 8822--8826.

\bibitem{press2020hedges}
W.~H. Press, J.~A. Hawkins, S.~K. Jones~Jr, J.~M. Schaub, and I.~J.
  Finkelstein, ``Hedges error-correcting code for {DNA} storage corrects indels
  and allows sequence constraints,'' \emph{Proceedings of the National Academy
  of Sciences}, vol. 117, no.~31, pp. 18\,489--18\,496, 2020.

\bibitem{maarouf2022concatenated}
I.~Maarouf, A.~Lenz, L.~Welter, A.~Wachter-Zeh, E.~Rosnes, and A.~G. i~Amat,
  ``Concatenated codes for multiple reads of a {DNA} sequence,'' \emph{IEEE
  Transactions on Information Theory}, vol.~69, no.~2, pp. 910--927, 2023.

\bibitem{welzel2023dna}
M.~Welzel, P.~M. Schwarz, H.~F. L{\"o}chel, T.~Kabdullayeva, S.~Clemens,
  A.~Becker, B.~Freisleben, and D.~Heider, ``{DNA}-aeon provides flexible
  arithmetic coding for constraint adherence and error correction in {DNA}
  storage,'' \emph{Nature Communications}, vol.~14, no.~1, p. 628, 2023.

\bibitem{10619614}
S.~{Kas Hanna}, ``Short systematic codes for correcting random edit errors in
  {DNA} storage,'' in \emph{2024 IEEE International Symposium on Information
  Theory (ISIT)}, 2024, pp. 663--668.

\bibitem{hanna2025gccodesystematicshort}
\BIBentryALTinterwordspacing
------, ``{GC+} code: A systematic short blocklength code for correcting random
  edit errors in {DNA} storage,'' 2025. [Online]. Available:
  \url{https://arxiv.org/abs/2402.01244}
\BIBentrySTDinterwordspacing

\bibitem{11154528}
R.~Khabbaz, M.~Antonini, and S.~{Kas Hanna}, ``Marker guess \& check plus
  ({MGC+}): An efficient short blocklength code for random edit errors,'' in
  \emph{2025 13th International Symposium on Topics in Coding (ISTC)}, 2025,
  pp. 1--5.

\bibitem{reed1960polynomial}
I.~S. Reed and G.~Solomon, ``Polynomial codes over certain finite fields,''
  \emph{Journal of the society for industrial and applied mathematics}, vol.~8,
  no.~2, pp. 300--304, 1960.

\bibitem{10750859}
D.~Bar-Lev, O.~Sabary, R.~Gabrys, and E.~Yaakobi, ``Cover your bases: How to
  minimize the sequencing coverage in {DNA} storage systems,'' \emph{IEEE
  Transactions on Information Theory}, vol.~71, no.~1, pp. 192--218, 2025.

\bibitem{preuss2024sequencing}
I.~Preuss, B.~Galili, Z.~Yakhini, and L.~Anavy, ``Sequencing coverage analysis
  for combinatorial {DNA}-based storage systems,'' \emph{IEEE Transactions on
  Molecular, Biological, and Multi-Scale Communications}, 2024.

\bibitem{sokolovskii2024coding}
R.~Sokolovskii, P.~Agarwal, L.~Alberto~Croquevielle, Z.~Zhou, and T.~Heinis,
  ``Coding over coupon collector channels for combinatorial motif-based {DNA}
  storage,'' \emph{IEEE Transactions on Communications}, vol.~73, no.~6, pp.
  3750--3760, 2025.

\bibitem{cohen2024optimizing}
T.~Cohen and E.~Yaakobi, ``Optimizing the decoding probability and coverage
  ratio of composite {DNA},'' in \emph{2024 IEEE International Symposium on
  Information Theory (ISIT)}.\hskip 1em plus 0.5em minus 0.4em\relax IEEE,
  2024, pp. 1949--1954.

\bibitem{preuss2024efficient}
I.~Preuss, M.~Rosenberg, Z.~Yakhini, and L.~Anavy, ``Efficient {DNA}-based data
  storage using shortmer combinatorial encoding,'' \emph{Scientific reports},
  vol.~14, no.~1, p. 7731, 2024.

\bibitem{abraham2024covering}
H.~Abraham, R.~Gabrys, and E.~Yaakobi, ``Covering all bases: The next inning in
  {DNA} sequencing efficiency,'' in \emph{2024 IEEE International Symposium on
  Information Theory (ISIT)}.\hskip 1em plus 0.5em minus 0.4em\relax IEEE,
  2024, pp. 464--469.

\bibitem{10619151}
A.~Gruica, D.~Bar-Lev, A.~Ravagnani, and E.~Yaakobi, ``A combinatorial
  perspective on random access efficiency for {DNA} storage,'' in \emph{2024
  IEEE International Symposium on Information Theory (ISIT)}, 2024, pp.
  675--680.

\bibitem{chandak2019improved}
S.~Chandak, K.~Tatwawadi, B.~Lau, J.~Mardia, M.~Kubit, J.~Neu, P.~Griffin,
  M.~Wootters, T.~Weissman, and H.~Ji, ``Improved read/write cost tradeoff in
  {DNA}-based data storage using ldpc codes,'' in \emph{2019 57th Annual
  Allerton Conference on Communication, Control, and Computing
  (Allerton)}.\hskip 1em plus 0.5em minus 0.4em\relax IEEE, 2019, pp. 147--156.

\bibitem{cheung1989more}
K.-M. Cheung, ``More on the decoder error probability for reed-solomon codes,''
  \emph{IEEE Transactions on Information Theory}, vol.~35, no.~4, pp. 895--900,
  1989.

\bibitem{berlekamp1984algebraic}
E.~R. Berlekamp, ``Algebraic coding theory, revised ed,'' \emph{Laguna Hills,
  CA: Aegean Park}, 1984.

\bibitem{chen2020quantifying}
Y.-J. Chen, C.~N. Takahashi, L.~Organick, C.~Bee, S.~D. Ang, P.~Weiss, B.~Peck,
  G.~Seelig, L.~Ceze, and K.~Strauss, ``Quantifying molecular bias in {DNA}
  data storage,'' \emph{Nature communications}, vol.~11, no.~1, p. 3264, 2020.

\bibitem{mceliece2003decoder}
R.~McEliece and L.~Swanson, ``On the decoder error probability for reed-solomon
  codes (corresp.),'' \emph{IEEE Transactions on Information Theory}, vol.~32,
  no.~5, pp. 701--703, 2003.

\bibitem{miao2025errorratebinarybch}
\BIBentryALTinterwordspacing
S.~Miao, J.~Mandelbaum, H.~Jäkel, and L.~Schmalen, ``On the error rate of
  binary bch codes under error-and-erasure decoding,'' 2025. [Online].
  Available: \url{https://arxiv.org/abs/2509.24794}
\BIBentrySTDinterwordspacing

\bibitem{cai2021correcting}
K.~Cai, Y.~M. Chee, R.~Gabrys, H.~M. Kiah, and T.~T. Nguyen, ``Correcting a
  single indel/edit for {DNA}-based data storage: Linear-time encoders and
  order-optimality,'' \emph{IEEE Transactions on Information Theory}, vol.~67,
  no.~6, pp. 3438--3451, 2021.

\bibitem{shevtsova2010improvement}
I.~G. Shevtsova, ``An improvement of convergence rate estimates in the lyapunov
  theorem.'' in \emph{Doklady Mathematics}, vol.~82, no.~3, 2010.

\end{thebibliography}


\end{document}